\documentclass[lettersize,journal]{IEEEtran}
\usepackage{array}
\usepackage[caption=false,font=normalsize,labelfont=sf,textfont=sf]{subfig}
\usepackage{textcomp}
\usepackage{stfloats}
\usepackage{url}
\usepackage{verbatim}
\usepackage{graphicx}
\usepackage{cite}
\usepackage{amsmath,amssymb,amsfonts}
\usepackage[caption=false,font=normalsize,labelfont=sf,textfont=sf]{subfig}
\usepackage[export]{adjustbox}
\usepackage{multirow,multicol}
\usepackage{enumitem}
\usepackage{colortbl}
\IEEEoverridecommandlockouts
\usepackage{changepage}
\usepackage{calc}
\usepackage{hyperref}
\hypersetup{
    colorlinks,
    citecolor=blue,
    filecolor=black,
    linkcolor=blue,
    urlcolor=black
}
\usepackage{algorithm}
\usepackage{algpseudocode}
\usepackage{pifont}
\usepackage[table]{xcolor}  
\usepackage{circledsteps}

\usepackage[para,online,flushleft]{threeparttable}
\usepackage{bm}
\usepackage{rotating}
\usepackage{listings}
\usepackage{parcolumns}
\usepackage{xcolor}

\algrenewcommand\algorithmicrequire{\textbf{Input:}}
\algrenewcommand\algorithmicensure{\textbf{Output:}}

\usepackage{caption}
\newcounter{heuristic}

\newcommand{\heuristiccaption}[1]{%
  \refstepcounter{heuristic}%
  \renewcommand{\thealgorithm}{\arabic{heuristic}}%
  \captionsetup{name=Heuristic}%
  \caption{#1}%
}

\newcommand{\pluseq}{\mathrel{+}=}

\usepackage{float} 

\newcommand{\shamim}[1]{{\color{blue} #1}}

\begin{document}

\title{Differentiation Between Faults and Cyberattacks through Combined Analysis of Cyberspace Logs and Physical Measurements}

\author{Mohammad Shamim Ahsan, Haizhou Wang, Venkateswara Reddy Motakatla, Minghui Zhu, and Peng Liu,~\IEEEmembership{Member,~IEEE}


\thanks{\shamim{This work has been published in the IEEE Transactions on Smart Grid, available at https://doi.org/10.1109/TSG.2026.3728276}}

\thanks{Mohammad Shamim Ahsan, Haizhou Wang and Peng Liu are with the College of Information Sciences and Technology, Pennsylvania State University, University Park, PA 16801 USA. (email: msa6152@psu.edu, hjw5074@psu.edu, pxl20@psu.edu)}

\thanks{Venkateswara Reddy Motakatla is with the National Renewable Energy Laboratory (NREL), Golden CO, USA. (email: VenkateswaraReddy.Motakatla@nrel.gov)}

\thanks{Minghui Zhu is with the Department of Electrical Engineering, The Pennsylvania State University, University Park, PA 16802 USA (email: muz16@psu.edu)}

}

\markboth{Journal of \LaTeX\ Class Files,~Vol.~XX, No.~X, September~2025}%
{Shell \MakeLowercase{\textit{et al.}}: A Sample Article Using IEEEtran.cls for IEEE Journals}


\maketitle

\begin{abstract}
In recent years, cyberattacks - along with physical faults - have become an increasing factor causing system failures, especially in DER (Distributed Energy Resources) systems. In addition, according to the literature, a number of faults have been reported to remain undetected. Consequently, unlike anomaly detection works that only identify abnormalities, differentiating undetected faults and cyberattacks is a challenging task. Although several works have studied this problem, they crucially fall short of achieving an accurate distinction due to the reliance on physical laws or physical measurements. To resolve this issue, the industry typically conducts an integrated analysis with physical measurements and cyberspace information. Nevertheless, this industry approach consumes a significant amount of time due to the manual efforts required in the analysis. In this work, we focus on addressing these crucial gaps by proposing a non-trivial approach of distinguishing undetected faults and cyberattacks in DER systems. Specifically, first, a special kind of dependency graph is constructed using a novel virtual physical variable-oriented taint analysis (PVOTA) algorithm. Then, the graph is simplified using an innovative node pruning technique, which is based on a set of context-dependent operations. Next, a set of patterns capturing domain-specific knowledge is derived to bridge the semantic gaps between the cyber and physical sides. Finally, these patterns are matched to the relevant events that occurred during failure incidents, and possible root causes are concluded based on the pattern matching results. In the end, the efficacy of the proposed semi-automatic integrated analysis is evaluated through five case studies covering failure incidents caused by FDI attacks, system faults, and memory corruption attacks, including a false-classification scenario.
\end{abstract}


\section{Introduction}\label{sec:intro}

\IEEEPARstart{D}{ifferentiation} between undetected faults and cyber attacks is a critical issue in DER (Distributed Energy Resources) systems, due to at least two main reasons. First, catastrophic failure incidents (e.g., outages) in modern power systems are no longer entirely caused by system faults: cyber attacks are an increasing factor. For example, (a) the complete blackout in Ukraine due to a large-scale cyber attack~\cite{inc2_liang20162015}; (b) the Iranian cyberattack on industrial PLCs located in many countries, including multiple US states~\cite{inc3_irgc2020}. Second, many faults in DER systems are {\em undetected faults}. For example, (a) standard protection devices like fuses and circuit breakers may not be sufficient to detect all types of faults in solar photovoltaic (PV) systems; (b) faults that occur under low irradiation conditions may go undetected; (c) PV arrays can experience multiple faults simultaneously, confusing the fault detectors. Third, if differentiation is ignored, the failure recovery procedure could be seriously affected. For example, system operators may mistakenly conclude that a failure is caused by faults when it is actually caused by a cyber attack. This often leads to ineffective incident response and recovery. Ultimately, the affected system becomes more vulnerable and less resilient~\cite{tertytchny2020classifying}. It is widely recognized in the literature that differentiation between undetected faults and cyber attacks is indispensable~\cite{syfert2022integrated}. 

A significant number of works in the literature~\cite{benninger2020anomaly, wang2022online, tsai2020anomaly} have focused on detecting anomalies (e.g., power outages). However, techniques for detecting anomalies and techniques for differentiating between faults and cyber attacks are, in principle, different. Specifically, these anomaly detection systems (ADSs) can detect only whether something is abnormal, not the specific type of anomaly. Several studies have been conducted on distinguishing undetected faults and cyberattacks~\cite{amini2017hierarchical, gupta2022distinguishing, sahoo2020resilient, ao2016adaptive, sakhnini2021physical, ganjkhani2021integrated, beg2021cyber, shen2024detection}. These works focus on the physical side of CPSs and are based on physical laws or physical measurements. Unfortunately, reliance on physical laws makes these works limited to effectively handle real-world signal noise~\cite{mo2011cyber}, and the reliance (e.g., take a machine learning approach) on physical measurements is prone to suffer from inaccuracy due to the imbalanced data in CPS~\cite{madabhushi2023survey} and from limited explainability. To mitigate these issues, the DER industry utilizes the analysis results of the existing detection and differentiation techniques but does not treat them as the final outcome. Rather, a team of human analysts is synthesizing the various intermediate analysis results to integrate the findings from not only physical side analyzers (e.g., ADSs, localization analyzers) but also cyberspace analyzers (e.g., data dependency analyzers). A main goal is to integrate physical and cyberspace information (e.g., obtained through taint analysis) to avoid severe consequences caused by incorrect cause identification. Specifically, cyberspace uniquely contains some critical information, especially a variety of logs and data dependencies that can facilitate the task of correctly differentiating undetected faults and various cyberattacks.

\begin{figure}[htbp]
    \centering \includegraphics[scale=0.6511]{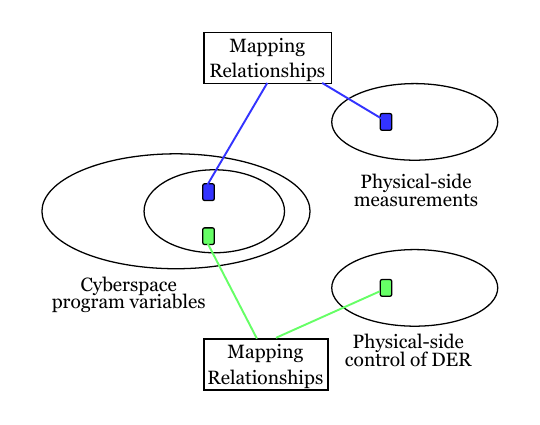}
    \caption{Mapping between cyberspace program variables and physical system variables in a DER system}
    \label{fig:insight}
\end{figure}

However, the industry approach is very time-consuming, as it involves a lot of manual effort. In this paper, we propose an approach to take the first step towards automatically synthesizing physical side and cyberspace findings. The mapping relationships shown in Figure~\ref{fig:insight} provide a fundamental reason why combined analysis could create a unique capability. In this figure, the mapping relationships on the top of the figure are between the virtual physical variables (from which physical side measurements are collected) and the program variables used by DERMS (DER Management System) applications, and the mapping relationships on the bottom of the figure are between the program variables used by DERMS applications and physical side control of DER. Note that the messages sent from DERMS applications to DER devices play an essential role in achieving control of DER. Specifically, if taint propagation (i.e., information flow from one entity to another) is observed in the cyberspace from a variable that is mapped to a physical side measurement to a variable that is mapped to a physical side control, then the extended cyber-physical taint propagation would provide a unique lens to examine not only the (varied) effects of different cyberattacks, but also the differences between undetected faults and cyberattacks. Although this kind of mapping is highly desirable, it is non-trivial to achieve due to the following challenge: existing automatic taint analysis techniques \cite{t4_arzt2014flowdroid, t5_sridharan2011f4f} fall short of meeting the synthesized analysis requirements of a DER system. On one hand, forward taint analysis explicitly requires both the taint source (i.e., a particular program variable that appears in a specific line of code) and the sink in advance. However, the incident report used by analysts, though it contains a detailed description of the detected anomalies, usually does not tell what the specific taint sources are. On the other hand, backward taint analysis identifies all the possible taint sources for a given sink, but most of the identified sources are usually irrelevant to the reported incident, and examining them can waste a considerable amount of resources. Moreover, currently, the connections between physical measurements and cyberspace information items (e.g., data dependencies between program variables) are not being automatically analyzed in real-world practice. In fact, significant semantic gaps exist between physical measurements and cyberspace information items. 

To address these challenges, we develop two new ideas. First, a virtual physical variable-oriented taint analysis (\textit{PVOTA}) algorithm is introduced, where the virtual physical variable(s) mentioned in the incident report is used to guide our analyzer to automatically find the corresponding tainted sources and taint propagation paths. Second, a set of patterns is designed, where each pattern is devised to help bridge the semantic gap between physical side measurements and cyberspace information. These two ideas make our proposed approach non-trivial. In particular, a special kind of dependency graph is constructed based on the first idea of \textit{PVOTA}. Specifically, we start with virtual physical variable(s): first, a backtracking mechanism along with a heuristic is employed to find the taint sources; second, a forward tracking mechanism is utilized to identify the taint propagation paths that reach the sink. Then, a set of context-dependent operations is performed to prune the less important nodes from the graph. Assisted by the constructed simplified dependency graph, a set of specific patterns is leveraged to cross-check the observed physical-side and cyberspace information items, and we audit the matched patterns. After that, the matching results are synthesized to conclude about the root cause of the failure incident. The proposed approach makes the differentiating task much more time-efficient than the current industry approach. Accordingly, manual efforts, and hence the required time, are substantially reduced. In summary, the following contributions are made in this work:
\begin{itemize}
    \item A virtual physical variable-oriented taint analysis (\textit{PVOTA}) technique is proposed to construct dependency graphs, especially in DER systems.
    \item A set of patterns is derived to bridge the semantic gap between the cyber and physical sides. Assisted by the dependency graph constructed by the PVOTA technique, we match these patterns against the relevant events that had happened during each reported incident, and a decision on the root cause is made by observing the pattern-matching results. 
    \item To evaluate the feasibility and efficacy of the proposed combined analysis approach, we conduct several case studies that cover three representative types of incidents: an incident caused by a False Data Injection attack, an incident caused by undetected faults, and an incident caused by a memory corruption attack. In addition, a false-classification scenario is explored to investigate the conditions under which the proposed approach may may lead to an incorrect root-cause identification.
\end{itemize}

\section{Background and Existing Works}\label{sec:bgm} 

\subsection{DER Systems and Cyberattacks}
Typically, a DERMS application is used by utilities to manage diverse and dispersed DERs, both individually and in aggregate. From the lowest level, different services of the DERs (e.g., individual EVs, rooftop PVs, household batteries, etc.) or microgrids are coordinated and scheduled by a flexible resource scheduler (FRS), which operates as a local DERMS for its substation. Both utility-scale and behind-the-meter DERs can be managed either directly by the FRS or through a transactive market manager (TMM) or DER aggregators. In either case, a DERMS application is used to dispatch DER resources and real-time control. The next two upper layers of a DERMS are the DSO (Distributed System Operator) and the TSO (Transmission System Operator), which coordinate and manage signals from all substations to provide distribution and transmission services. 

The distributed nature and complex interconnectivity of DERMS make it vulnerable to cyberattacks, particularly false data injection (FDI), memory corruption, and man-in-the-middle (MITM) attacks. In an FDI attack, the attacker tries to manipulate DER operations by injecting falsified data of various types. Considering the distributed architecture of DERMS, any erroneous data injected into the communication layer can result in a cascading effect on the DERs that affects grid stability, control operations, and normal voltage or current flows. In an MITM attack, the attacker intercepts or modifies critical data by compromising a gateway, a router, communication channels, or other non-end equipment. During a memory corruption attack, the attacker exploits the vulnerability of DERMS application program executions by overwriting or overflowing different memory segments, notably the stack and heap.

\subsection{Undetected Faults in PV Systems}
In photovoltaic (PV) systems, ground faults, arc faults, and line-to-line faults are identified as major causes of catastrophic failures~\cite{alam2015comprehensive}. Regarding protection devices, ground fault detection and interruption fuse (GFDI) protects equipment from being affected by a significant current flow to the ground circuit caused by an accidental connection between a current-carrying conductor (CCC) and an equipment grounding conductor (EGC) or earth~\cite{alam2015comprehensive}. However, the leakage current can return to the PV conduction path through these fuses, which is an unintentional systematic drawback of this device. Arc-fault circuit interrupter (AFCI) is another protection device that protects PV arrays from being fire hazards caused due to the discontinuities in the CCCs or insulation breakdown in adjacent CCCs. To handle line-to-line faults, overcurrent protection devices (OCPDs) are used when the fault current is higher than the rated current of OCPD, which is at most 1.56 times the short-circuit current~\cite{zhao2011fault}. However, these faults may remain undetected when they occur under low irradiance since the current is not large enough to melt the OPCD~\cite{zhao2011fault}. 

\subsection{Taint Analysis}
Taint analysis is a program analysis technique that tracks the information flow originating from untrusted sources (e.g., API functions, user inputs), which propagates along the program execution path and eventually reaches release points or sensitive functions (known as \textit{sinks}). Data entered into the program by these sources is considered \textit{tainted}, thereby causing all dependent variables to be tainted as well. An example of such propagation can be described using the code statements: \lstinline[basicstyle=\small\ttfamily]|x=get_input(); y=1; z=x; w=y+z;|, where \texttt{get\_input()} is a taint source, as untrusted data can enter the program through it. Consequently, \texttt{x} becomes tainted. However, \texttt{y} is untainted as it allocates a constant value. Since \texttt{z} has data dependency on \texttt{x}, it becomes tainted as well. According to the taint propagation rule: \textit{tainted}+\textit{untainted}=\textit{tainted}, \texttt{w} is also considered tainted. Due to these data dependencies in programs, an attacker can exploit taint sources to manipulate inputs and critical computations, and taint analysis can effectively trace the propagation paths of such tainted data through the program. 
\subsection{Related Works}
Although works on differentiating faults and attacks are comparatively less than that of detecting, existing research can be categorized into two groups: physical law-based and physical side measurements-driven.

In the first group, approaches depend on the physical laws or mathematical foundations of system design and typically make a set of assumptions prior to modeling. Optimization-based frequency-domain approach~\cite{amini2017hierarchical}, mapping of physics law-driven trajectories on a frequency-active voltage-centric Cartesian plane~\cite{gupta2022distinguishing}, asymmetric relationship-based approach for analyzing compromised frequency signals~\cite{sahoo2020resilient}, residual signals construction using an adaptive sliding mode observer technique~\cite{ao2016adaptive} are the most prominent systems in the literature. Regarding physical side measurement-driven approaches, they predict based on real-time and historical power measurements. Sakhini et al.~\cite{sakhnini2021physical} classify attacks and faults with localization to particular measurements in the operational layer (a.k.a. physical layer) of smart grids using an ensemble algorithm and data representation learning. A deep neural network (DNN)-based strategy is proposed in~\cite{ganjkhani2021integrated} using historical outage data received from protection relays (both feeder and DER) and fault indicators to classify physical faults and different cyberattacks, including FDI and replay attacks. Other well-known strategies utilize parametric time-frequency logic (PTFL)~\cite{beg2021cyber}, multi-label CNN with matrix singular value decomposition (SVD)~\cite{shen2024detection}, etc.
\\ \\\noindent{\textbf{Limitations.}} The primary limitation of physical law-based approaches is the inefficiency of handling real-world signal noise~\cite{mo2011cyber}. In addition, these models are often constrained by a predefined set of assumptions (e.g., losslessness of the system, dependency on frequency-specific synchronized PMU measurements), which may limit their applicability in some situations. With regard to approaches that rely on physical side measurements, the fundamentally crucial issue is the quality of the train-test datasets. In most cases, an imbalanced data distribution is observed in various fault and attack classes, severely affecting these approaches, especially supervised machine learning-based models~\cite{madabhushi2023survey}. One convincing reason behind this is the scarcity of anomalous data, since system failure incidents are not particularly frequent. Moreover, these physical measurement-driven approaches are also difficult to explain. 

\subsection{Key Differences from Prior Works}
To address the limitations of existing approaches, we propose a semi-automatic integrated analysis that leverages cyberspace logs and physical measurements through taint-styled dependency graphs and pattern matching to differentiate undetected faults and cyberattacks. A comprehensive comparison between existing approaches and the proposed combined analysis is presented in Table~\ref{tab:comparison}. The comparison spans seven characteristics, including reliance on physical-system assumptions, sensitivity to class imbalance, robustness to signal and measurement noise, the utilization of physical measurements, cyber-physical analysis capability, data dependency analysis, and explainability.
\begin{table}[htbp]
\caption{Key Differences from Prior Works}
\label{tab:comparison}
\begin{adjustwidth}{-0.5cm}{}
\def\arraystretch{1}
\resizebox{0.53\textwidth}{!}{
\begin{tabular}{lccc}
\hline
\shortstack{\textbf{Characteristic}\\\textbf{}} &
\shortstack{\textbf{Law-}\\\textbf{based}} &
\shortstack{\textbf{Meas.-}\\\textbf{driven}} &
\shortstack{\textbf{Proposed}\\\textbf{Approach}} \\
\hline
Reliance on physical-system assumptions & \checkmark & $\times$ & $\times$ \\
Sensitivity to class imbalance & n/a & \checkmark & $\times$ \\
Noise robustness & Low & Moderate & High \\
Cyber-physical analysis & $\times$ & $\times$ & \checkmark \\
Data dependency analysis & $\times$ & $\times$ & \checkmark \\
Physical-side measurements utilization & \checkmark & \checkmark & \checkmark \\
Explainability & Partial & Partial & Full \\
\hline
\end{tabular}}
\end{adjustwidth}
\end{table}

\section{Proposed Approach}\label{sec:appr}

\subsection{Problem Statement}
How to semi-automatically conduct a combined analysis to achieve differentiation between undetected faults and cyber-attacks? By combined analysis, we mean the following: if taint propagation is observed in the cyberspace from a DERMS application program variable that is mapped to a physical side measurement to a variable that is mapped to a physical side control, then the extended cyber-physical taint propagation would provide a unique lens to examine the differences between undetected faults and cyber-attacks. It should be noticed that the techniques for detecting anomalies and the techniques for differentiating between faults and cyber-attacks are {\em inherently} different. An ADS detects anomalies, but does not differentiate between the {\em causes} for the detected anomaly.
\\\\
\noindent{\textbf{Definitions.}} We define a new term, \textit{virtual physical variable}, which represents a DERMS application program variable that is used to hold the value of physical side measurements. In addition, the set of virtual physical variables, taint sources, and sinks are represented by $P_{var}$, $T_{src}$, and $T_{snk}$, respectively. Moreover, a variable is considered as an \textit{auxiliary variable}, denoted as $V_s$, if it is neither a virtual physical variable, a function call, nor an array/object related variable.

\subsection{Overview}\label{sec:overview}
\begin{figure*}[!ht]
    \begin{adjustwidth}{-0.35cm}{} 
        \includegraphics[scale=0.37]{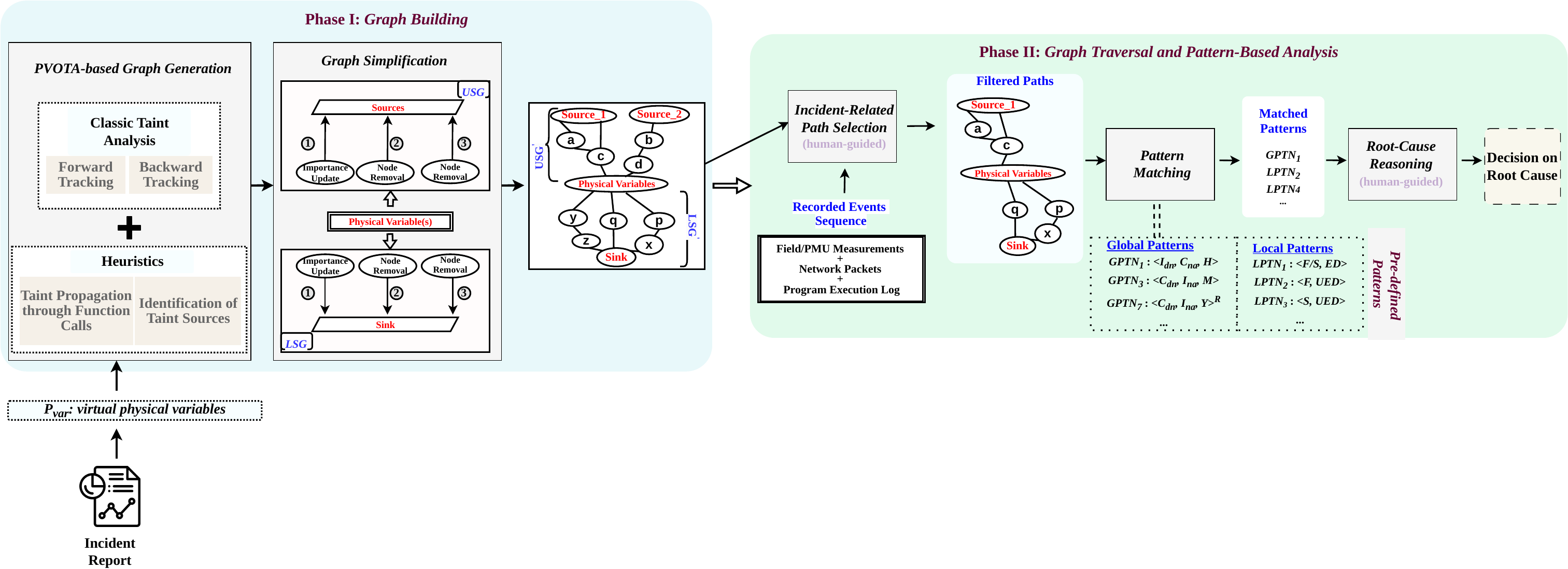}
    \end{adjustwidth}
    \vspace{-15pt}
    \caption{Overview of the proposed approach}
    \label{fig:overview}
\end{figure*}

In this paper, we propose a novel approach to distinguish undetected faults and cyberattacks utilizing both physical-side measurements and cyberspace information. In particular, a virtual physical variable-oriented taint analysis technique, \textit{PVOTA}, is introduced to generate dependency graphs, and a set of domain-specific patterns is derived from physical-side measurements and cyberspace information to capture the distinct characteristics of individual causes. Specifically, the proposed approach consists of two main phases: (i) graph building, and (ii) graph traversal and pattern-based analysis, as illustrated in Figure~\ref{fig:overview}. In the graph-building phase, virtual physical variable(s) are first collected from the incident report. Next, two dependency subgraphs—the upper subgraph (\textit{USG}) and lower subgraph (\textit{LSG})—are constructed based on these variables using backward and forward taint tracking, respectively. Subsequently, a graph simplification algorithm is applied to remove less important nodes from the subgraphs using a set of context-dependent operations. At the beginning of the second phase, incident-related paths in the \textit{USG} and \textit{LSG} are selected based on a sequence of events recorded during the system failure. Then, each node along these paths is traversed to identify matched patterns, which are subsequently used to determine the root cause. In the following two subsections, each phase is described in detail.
\\ \\
\subsection{Phase I: Graph Building} 
This is the first phase of our proposed approach that comprises two steps: graph generation and graph simplification. In the first stage, a specialized dependency graph is constructed using a non-trivial taint analysis technique. In the subsequent stage, a novel graph pruning strategy is employed to retain only the most relevant nodes in the graph.

\subsubsection{Graph Generation Technique}\label{sec:generation}
\begin{figure}[htbp]
    \centering 
        \includegraphics[scale=0.59]{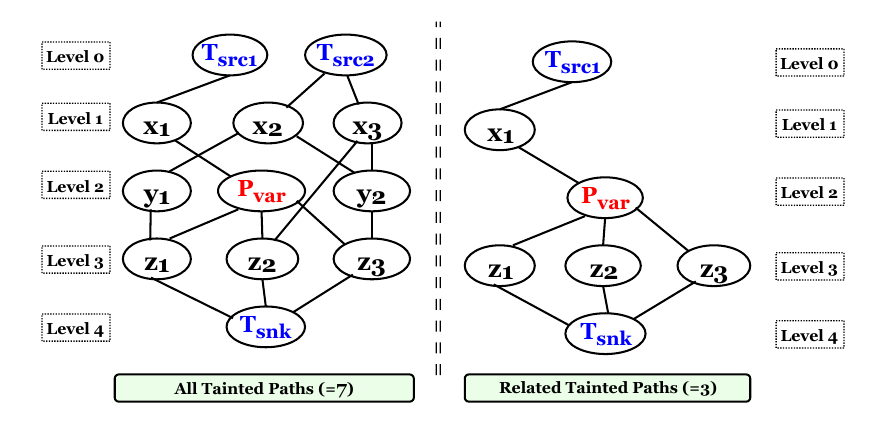}
    \caption{An example to demonstrate the shortcomings of the traditional taint analysis technique}
    \label{fig:valid}
\end{figure}

To investigate the root cause of system failures based on both physical-side measurements and cyberspace information, such as server application code, network packets, and DERMS application logs, it is arguably essential to identify entry points, precise propagation paths, and the exit point that has affected related DER devices. This observation initiates the necessity of employing a taint analysis technique, where taint propagation is observed from a given source to the distinguished sink. However, standard taint analysis is not appropriate, especially for DER systems. One of the major issues is that not all generated tainted paths are related to causing system failure; only those that include virtual physical variables are relevant since real-world is affected by the alternation of their measurements. For example, for the graph illustrated in Figure~\ref{fig:valid}, almost half of the tainted paths generated by a standard taint propagation technique are irrelevant to the root cause analysis, specifically for the cyber-physical system (CPS) domain. Another shortcoming of standard taint analysis is the requirement of having prior knowledge of the sources and sinks, which is not always feasible. In contrast, our proposed approach adopts a taint analysis method and overcomes these limitations by utilizing a forward and backward propagating strategy to generate tainted paths started from virtual physical variables, along with a heuristic to determine the potential tainted sources. 
\\

\noindent{\textit{Approach.}} The proposed dependency graph construction approach enhances standard backward and forward taint analysis techniques using non-trivial virtual physical variable-oriented strategy, along with two heuristics (see Algorithm~\ref{alg:graph_construction}). Each node in the graph contains the specific tainted entity in the program. A tainted entity can be either an identifier (e.g., variable, object, referencing, expression) or a function call. On the other hand, similar to the classic taint propagation technique, our proposed technique leverages directed edges that represent data dependency. Specifically, an edge from node $a$ to node $b$ is denoted as $E(a,b)$. During forward traversal, this edge indicates the propagation of the taint from $a$ to $b$. However, while backtracking, the tainted data flows in the opposite direction, from $b$ to $a$ for the edge $E(a,b)$.

At the beginning of the graph construction, we start with $P_{var_i}\in P_{var}$ and propagate upward using a backtracking approach to generate {\it{Upper Subgraph (USG)}}, as described in lines 1--8 of Algorithm~\ref{alg:graph_construction}. Next, a forward traversal technique is applied to construct {\it{Lower Subgraph (LSG)}}, which begins at $P_{var_i}$ and ends whenever any $T_{snk_i} \in T_{snk}$ is found. Lines 9--18 of Algorithm~\ref{alg:graph_construction} show the construction of LSG. The back-front traversal is similar to the standard taint analysis technique. However, for taint propagation through functions, we introduce several modifications, as described in Heuristic~\ref{alg:heuristics_2} and illustrated in Figures~\ref{fig:gen_example} and~\ref{fig:inter_flow} (referring to the Appendix). In addition, to avoid complications, we ignore the cases where a traversing function is called by another function (i.e., an indirect connection as shown in Figure~\ref{fig:ind_conn}). Since during backtracking, the taint sources and their volume are unknown, a heuristic is proposed to find potential taint sources (see Heuristic~\ref{alg:heuristics_1} in Appendix). Specifically, two API calls: \texttt{GridAPPSD(username, password)} and \texttt{conn.get\_response(t.TIMESERIES, message)} are identified as tainted sources during the case studies conducted later. At the beginning of the heuristic, two observations are made on potential candidates of the taint sources, which are: (i) user-defined functions cannot be taint sources, and (ii) potential taint sources typically include: API calls, command-line argument handlers, query-processing functions, and read functions. Then, a node-weighting strategy is employed in the generated {\it{USG}} to find the set of taint sources ($T_{src}$). 
\begin{algorithm}[t]
\footnotesize
\caption{Graph Construction using PVOTA}
\label{alg:graph_construction}

\begin{algorithmic}
\Require $P_{var}$: Virtual Physical Variables
\Ensure $USG, LSG$: Dependency Graphs 
\end{algorithmic}

\begin{algorithmic}[1]

\Statex {\color{lightgray} // \textit{Step 1: Construct the Upper Subgraph ($USG$)}}
\ForAll{$P_{var_i} \in P_{var}$}

    \State Initialize $USG \gets \emptyset$
    \State Start backtracking traversal from $P_{var_i}$.
    \State For each edge $E(a,b)$, propagate taint backward from $b$ to $a$.
    \State Apply the backtracking and inter-procedural rules in Heuristic~\ref{alg:heuristics_2}.
    \State Identify potential taint sources using Heuristic~\ref{alg:heuristics_1}.
    \State Add all visited nodes and edges to $USG$.

\EndFor

\Statex {\color{lightgray} // \textit{Step 2: Construct the Lower Subgraph ($LSG$)}}
\ForAll{$P_{var_i} \in P_{var}$}

    \State Initialize $LSG \gets \emptyset$
    \State Start forward traversal from $P_{var_i}$.
    \State For each edge $E(a,b)$, propagate taint forward from $a$ to $b$.
    \State Apply the forward-tracking, callee-propagation, and callback rules in Heuristic~\ref{alg:heuristics_2}.
    \State Add all visited nodes and edges to $LSG$.

    \If{a taint sink $T_{snk_i} \in T_{snk}$ is reached}
        \State Stop forward traversal.
    \EndIf

\EndFor

\State \Return $USG, LSG$

\end{algorithmic}
\end{algorithm}

\subsubsection{Graph Simplification Technique}\label{sec:simplification}
To increase the readability and usability of the generated graphs, we develop a context-aware graph simplification algorithm, where less significant nodes are eliminated using several context-based operations. Specifically, we develop two ideas: i) make some nodes more important than others, and ii) the importance of a node depends on contextual situations. To tackle the first aspect, we assign ``importance value" to each node, denoted as $imv_k(N_k)$, where $N_k$ represents the $k^{th}$ node. Besides, $\Delta imv_k(N_k)$ represents the change (usually, increase) in the importance value for node $N_k$. However, for simplicity, $imv_k$ is often used in this work to denote the importance value of a node, in general. Moreover, several contextual situations are defined, such as single and multi-dimensional arrays or object referencing, inner referencing, functions, expressions, etc. Specifically, two different node reduction techniques are devised to optimize the subgraphs {\it{USG}} and {\it{LSG}} separately. Each approach consists of two operations: \textit{importance update} and \textit{node removal}. An overview of these operations is illustrated in Figure~\ref{fig:overview}.

Whenever a subgraph is traversed, we call it a \textit{scan} during which either of the operations is performed. For example, in {\it{USG}}, scanning means traversing from the $T_{src_i}$ to $P_{var_i}$ once; while in {\it{LSG}}, it refers to visiting from the $P_{var_i}$ to $T_{snk_i}$. The overall simplification procedures for the upper and lower subgraphs are summarized in Algorithms~\ref{alg:usg_simplification} and~\ref{alg:lsg_simplification}, respectively. The proposed optimization techniques are driven by a set of contextual rules. Specifically, both techniques begin with an importance-update scan, during which node importance values are adjusted according to the contextual characteristics of the traversed nodes. Subsequently, one or more node-removal scans prune low-importance and redundant intermediate nodes while preserving critical propagation paths between the corresponding graph endpoints. The importance update rules are described in Heuristics~\ref{alg:heuristics_3} and~\ref{alg:heuristics_4}, including related illustrations in Figures~\ref{fig:usg_upd_example} and~\ref{fig:lsg_upd_example}. The resulting simplified subgraphs retain the most significant dependencies for subsequent analysis. For instance, in an expression assignment, the destination variable and the corresponding expression node are prioritized over operand nodes. Similarly, when array or object references are involved, referencing nodes are favored over intermediate reference nodes to facilitate subsequent graph reduction. 

Notably, the change in the importance value at each node, $\Delta imv_k(N_k)=1 \text{ or } \alpha$, where $\alpha$ is the ``importance factor". It is defined as $\alpha=\frac{1}{m}$, where $m$ is the minimum number of incoming edges with which a node is always preserved during the node removal operation. In this work, we take $m=2$, hence $\alpha=\frac{1}2{}$ so that a node with at least two incoming edges is never removed. \\
\begin{algorithm}[t]
\footnotesize
\caption{Upper Subgraph Simplification}
\label{alg:usg_simplification}

\begin{algorithmic}
\Require Upper Subgraph $USG$, virtual physical variables $P_{var}$, taint sources $T_{src}$
\Ensure Simplified Upper Subgraph $USG'$
\end{algorithmic}

\begin{algorithmic}[1]

\State Initialize $USG' \gets USG$.
\State Set $\alpha \gets \frac{1}{m}$
\State Initialize $imv_k(N_k) \gets 0$ for each node $N_k \in USG'$.
\State Set $imv_k(N_k) \gets 1$ for each $N_k \in P_{var} \cup T_{src}$.

\Statex {\color{lightgray}// \textit{Scan 1: Importance Update}}
\ForAll{paths from $P_{var_i}$ to $T_{src_i}$ in $USG'$}
    \State Traverse the path from $P_{var_i}$ to $T_{src_i}$.
    \ForAll{nodes $N_k$ on the path}
        \State Update $imv_k(N_k)$ using the USG importance-update rules in Heuristic~\ref{alg:heuristics_3}
    \EndFor
\EndFor

\Statex {\color{lightgray}// \textit{Scan 2: Low-Importance Node Removal}}
\ForAll{nodes $N_k \in USG'$}
    \If{$imv_k(N_k) < 1$}
        \State Remove $N_k$ from $USG'$.
        \State Reconnect incoming edges of $N_k$ to its next available successor.
    \EndIf
\EndFor

\Statex {\color{lightgray}// \textit{Scan 3: Intermediate Node Removal}}
\ForAll{three consecutive nodes $(N_{k-1}, N_k, N_{k+1})$ in $USG'$}
    \If{$imv_{k-1}=imv_k=imv_{k+1}=1$ \textbf{and} $N_k \in V_s$}
        \State Remove $N_k$ from $USG'$.
        \State Reconnect $N_{k-1}$ to $N_{k+1}$.
    \EndIf
\EndFor

\State \Return $USG'$

\end{algorithmic}
\end{algorithm}
\begin{algorithm}[t]
\footnotesize
\caption{Lower Subgraph Simplification}
\label{alg:lsg_simplification}
\begin{algorithmic}[1]

\Require Lower Subgraph $LSG$, virtual physical variables $P_{var}$, taint sinks $T_{snk}$
\Ensure Simplified Lower Subgraph $LSG'$

\State Initialize $LSG' \gets LSG$.
\State Set $\alpha \gets \frac{1}{m}$

\ForAll{nodes $N_k \in LSG'$}
    \State Initialize $imv(N_k)$ $\gets$ number of incoming edges.
\EndFor

\State Keep $imv(P_{var_i}) \gets 1$ and $imv(T_{snk_i}) \gets 1$.

\Statex {\color{lightgray}// \textit{Scan 1: Importance Update}}
\ForAll{forward-tracking paths in $LSG'$}
    \State Traverse the path from $P_{var_i}$ toward $T_{snk_i}$.
    \ForAll{nodes $N_k$ on the path}
        \State Update $imv(N_k)$ using Heuristic~\ref{alg:heuristics_4}.
    \EndFor
\EndFor

\Statex {\color{lightgray}// \textit{Scan 2: First Node Removal}}
\ForAll{nodes $N_k \in LSG'$}
    \If{$imv(N_k) \notin \mathbb{Z}^{+}$}
        \State Remove $N_k$ from $LSG'$.
        \If{$N_k$ has multiple outgoing edges}
            \State Redirect incoming edges of $N_k$ to the next available node.
        \EndIf
    \EndIf

    \If{$imv(N_{k+1}) - imv(N_k) = \alpha$ and $imv(N_{k+1}) \neq imv(N_{k+2})$}
        \State Remove $N_k$ from $LSG'$.
    \EndIf

    \If{$imv(N_{k+1}) = imv(N_k)$ and $imv(N_k) \notin \mathbb{Z}^{+}$}
        \State Remove $N_k$ from $LSG'$.
    \EndIf
\EndFor

\Statex {\color{lightgray}// \textit{Scan 3: Intermediate Node Removal}}
\ForAll{three consecutive nodes $(N_{k-1}, N_k, N_{k+1})$ in $LSG'$}
    \If{$imv(N_{k-1}) = 1$, $imv(N_k) = 1$, $imv(N_{k+1}) = 1$ and $N_k \in V_s$}
        \State Remove $N_k$ from $LSG'$.
        \State Reconnect $N_{k-1}$ to $N_{k+1}$.
    \EndIf
\EndFor

\State \Return $LSG'$

\end{algorithmic}
\end{algorithm}

\subsection{Phase II: Graph Traversal and Pattern-Based Analysis} \label{sec:patterns}
This phase consists of three stages: (i) Incident-Related Path Selection, (ii) Pattern Matching, and (iii) Root-Cause Reasoning. In general, taint propagates through a sequence of events, which are recorded along the tainted paths of the dependency graph. However, not all recorded events are necessarily related to the root cause of the incident. Therefore, this phase begins by selecting incident-related paths from the dependency graph constructed in Phase I. Specifically, a human analyst filters out unrelated paths using the sequence of events observed during the incident. Examples of such event sequences are presented in Section~\ref{sec:case_studies}. It should be noted that these sequences are incident-specific, and the identification of relevant paths relies on the manual inspection performed by the analyst. In particular, the analyst uses years of experience to recognize plausible discrepancies in event instances from normal behavior. Based on these observations, the affected nodes are traced in the dependency graph, thereby identifying the incident-related paths. In the next stage, each node in the filtered path(s) is traversed, and the collected cyber and physical logs are examined to identify matches with the predefined patterns described in Section~\ref{sec:patterns}. Finally, the matched patterns and their propagation characteristics are analyzed to infer the most likely root cause of the incident based on expert judgment. Algorithm~\ref{alg:pattern_matching} presents the detailed procedure for the automatic pattern-matching stage. In contrast, incident-related path selection (first stage) and root-cause reasoning (third stage) are analyst-guided activities that rely on recorded event sequences and expert judgment.
\begin{algorithm}[t]
\footnotesize
\caption{Pattern Matching on Incident-Related Paths}
\label{alg:pattern_matching}
\begin{algorithmic}[1]

\Require Incident-related paths $P_{rel}$, recorded event sequence $E_{seq}$, benign logs $L_b$, global patterns $GPTN$, local patterns $LPTN$
\Ensure Matched patterns $M$

\State Initialize $M \gets \emptyset$.
\State Construct benign histograms from $L_b$ for float-type variables.
\State Extract device, network, and application tuples required for global pattern matching from $E_{seq}$.

\Statex {\color{lightgray}// \textit{Global Pattern Matching}}
\ForAll{response or dispatch-related variables $x$}
    \State Compare $v_x^d$ and $v_x^n$, and determine $Idn$/$Cdn$.
    \State Compare $v_x^n$ and $v_x^a$, and determine $Ina$/$Cna$.
    \State Determine $Y$ or $N$ using the benign histogram of $x$.
    \State Determine the global pattern $GPTN_j$ corresponding to $\langle Idn/Cdn, Ina/Cna, Y/N\rangle$.
    \State $M \gets M \cup \{GPTN_j\}$.
\EndFor

\Statex {\color{lightgray}// \textit{Local Pattern Matching}}
\ForAll{paths $p \in P_{rel}$}
    \ForAll{nodes $N_k \in p$}
        \State Determine the node type of $N_k$: float, string, or others.
        \If{$N_k$ is others}
            \State $L \gets LPTN_5$.
            \State $M \gets M \cup \{L\}$.
            \State \textbf{continue}
        \EndIf
        \State Retrieve or estimate the observed value $v_k$ of $N_k$ from recorded event instances.
        \If{the value of $N_k$ is unavailable}
            \State $L \gets LPTN_4$
        \ElsIf{$N_k$ is float-type}
            \State Estimate deviation $\Delta_k$ of $v_k$ using benign histograms.
            \State Determine whether the observed deviation is \emph{expected}.
            \If{the observed deviation is \emph{expected}}
                \State $L \gets LPTN_1$
            \Else
                \State $L \gets LPTN_2$
            \EndIf
        \ElsIf{$N_k$ is string-type}
             \State Determine whether $v_k$ differs from its expected benign value.
            \If{$v_k$ is consistent with its expected benign value}
                \State $L \gets LPTN_1$
            \Else
                \State $L \gets LPTN_3$
            \EndIf
        \EndIf
        \State $M \gets M \cup \{L\}$.

    \EndFor
\EndFor

\State \Return $M$

\end{algorithmic}
\end{algorithm}
\\ \\
\noindent \textbf{Patterns.} In our proposed approach, the predefined patterns utilize three types of information: (i) data files holding PMU measurements (referred to as field measurements), (ii) Network packets PCAP (packet capture) files used to monitor network activities, and (iii) DERMS application log generated during the execution of the application program. Each variable in these logs is represented as a tuple: $\langle t, v\rangle$, where $t$ and $v$ denote the timestamp and value of the variable. In addition, $\langle t^d_x, v^d_x\rangle$, $\langle t^n_x, v^n_x\rangle$, and $\langle t^a_x, v^a_x\rangle$ are used to represent tuples for variable `$x$' in the corresponding logs. In a DER system, a dependency graph typically consists of three types of nodes: response message-related, dispatch message-related, and other program variables-related. However, for only response and dispatch messages, corresponding field measurements and network packets exist. In accordance with this knowledge, a set of patterns, referred to as \textit{global patterns} is devised to match abnormal system-wide event sequences. On the other hand, for other program variables that are only recorded in the application log, a separate set of patterns, called \textit{local patterns}, is used to track their values. Although global patterns are explicitly based on the time order, the constructed graph implicitly carries the time order of the executed program.
\\ \\
\noindent \underline{Global Patterns.} These patterns are developed to analyze the response and dispatch-related information available in every log. To monitor consistency between the log files, four new notations are introduced. In particular, $I_{dn}$ and $C_{dn}$ indicate inconsistency ($v^d_x \ne v^n_x$) and consistency ($v^d_x = v^n_x$), respectively, between the field measurements and the network packet. Likewise, $I_{na}$ and $C_{na}$ indicate inconsistency ($v^n_x \ne v^a_x$) and consistency ($v^n_x = v^a_x$), respectively, between the network packet and the application log. In addition, the orders of timestamps are $t^d_x < t^n_x < t^a_x$ and $t^d_x > t^n_x > t^a_x$ for response and dispatch, respectively. Here, two different time-orderings indicate the opposite directions of data flows during response (physical side $\rightarrow$ cyberspace) and dispatching (cyberspace $\rightarrow$ physical side). Typically, the value of any information, irrespective of a field measurement, network packet field, or program variable, can be a float or a string, depending on its content. These observed values are compared with the normal (`benign') ones to measure the \textit{deviation}. For a float-type variable, first, a histogram is plotted using the historical `benign' values of that variable. Let the total number of measurements and the total number of bins be $n$ and $m$, respectively. Moreover, the bins are $B_1, B_2, B_3, ..., B_m$ and the corresponding number of measurements in these bins are $n_1, n_2, n_3, ..., n_m$. Now, the value is considered to be \textit{deviated} if it cannot be located in any bin or is located in one of the smaller bins (say, bin $B_i$ and the relative size $p_i={n_i}/{n}$ is too small). However, for string-type variables, such statistical analysis is not needed since their values usually remain fixed in `benign' incidents. For either case, the existence and non-existence of deviation is denoted as $Y$ and $N$, respectively. Based on this strategy, each global pattern is represented as $GPTN_k\equiv \langle I_{dn}/C_{dn}, I_{na}/C_{na}, Y/N\rangle$. Eight such useful patterns are listed in Table~\ref{tab:new_patterns}. 
\\ \\
\noindent \underline{Local Patterns.} These patterns enable us to monitor potential deviations in each node of the graph using the DERMS application log. To represent the degree of deviation, a new notation $\Delta$ (Delta) is introduced. For a float-type node, $\Delta$ can be low (L), moderate (M), or high (H); whereas, for a string-type node, it is always considered high (H). The decision on $\Delta$ can be taken using the same histogram approach described in the global patterns section. Specifically, $\Delta$ is inversely proportional to $p_i$. The lower the $p_i$, the higher the likelihood of being considered as H; and vice versa. In any case, if the observed value cannot be assigned to any bin, then $\Delta$ is considered H. Here, this measurement is intentionally kept as abstract (i.e., low, moderate, and high) to support a customization facility for the security analyst. In addition, the degree of deviation is also system-dependent. For further analysis, two new concepts: \textit{expected deviation} and \textit{unexpected deviation} are defined, which represent what kind of behavior is expected to be observed at every node in the graph. If the degree of deviation at node \textit{X} is $\Delta_X$ and node \textit{Y} is a dependent of \textit{X}, then the expected deviation in \textit{Y} is referred to as $\Delta_X \rightarrow \Delta_Y$ and the unexpected deviation is denoted by $\Delta_X \rightarrow \Delta^{\prime}_Y$. The presence of $\Delta^{\prime}_Y$ emphasizes the possibility of a memory corruption attack in that node or any of its predecessor nodes. However, the existence of $\Delta_Y$ does not necessarily indicate any root cause; rather, it implies the need for continued analysis. Ideally, the `benign' values related to node $Y$ are available, and $\Delta_Y$ can be calculated using the proposed histogram approach. However, in practice, only a set of program variables is logged into the system. As a consequence, the values of $Y$ may not be available. In such cases, the degree of deviation can be inferred using $\Delta_X$ and/or the (observed and historical) values of \textit{X}. Since the response object contains requested measurements, such estimation is arguably feasible considering the contents of the node (e.g., array/object referencing, mathematical expression, etc.) and the concept of tainted data propagation implemented in our generated graph.

According to these observations and representations, we develop a set of local patterns listed in Table~\ref{tab:new_patterns}. Each pattern is formatted as $LPTN_k\equiv \langle F/S/Others, ED/UED/\top \rangle$, where $F$, $S$, and $Others$ represent the float-type node, string-type node, and others node (e.g., function call, object), respectively. In addition, the expected deviation and the unexpected deviation are denoted as $ED$ and $UED$, respectively. Moreover, in a few cases, the degree of deviation may not exist (considering different variants of code structures) and may not be applicable (function call). To handle such nodes, we use a new notation $\top$ (top). 
\begin{table}[htbp]
    \centering
    \caption{List of patterns. $GPTN_1 - GPTN_8$ are the global patterns and $LPTN_1 - LPTN_5$ are the local patterns. [$Y\equiv$deviation exists, $N\equiv$ deviation does not exist, $F\equiv$float, $S\equiv$string, $ED\equiv$expected deviation, $UED\equiv$unexpected deviation, $\top \equiv$unknown, $R\equiv$ dispatch-specific global patterns]}
    \begin{adjustwidth}{-0.5cm}{}
    \def\arraystretch{1}
    \resizebox{0.5311\textwidth}{!}{
    \begin{tabular}{lcc}
    \hline \textbf{Pattern \#} & \textbf{Definitions} & \textbf{Interpretation} \\\hline
    $GPTN_1$ & $\langle I_{dn}, C_{na}, Y \rangle$ & inconsistent network packet, deviation exists \\
    $GPTN_2$ & $\langle I_{dn}, C_{na}, N \rangle$ & inconsistent network packet, deviation does not exist \\
    $GPTN_3$ & $\langle C_{dn}, I_{na}, Y \rangle$ & inconsistent application log, deviation exists \\
    $GPTN_4$ & $\langle C_{dn}, I_{na}, N \rangle$ & inconsistent application log, deviation does not exist \\
    $GPTN_5$ & $\langle C_{dn}, C_{na}, Y \rangle$ & logs are consistent, deviation exists in field measurements\\
    $GPTN_6$ & $\langle C_{dn}, C_{na}, N \rangle$ & logs are consistent, no deviation in field measurements \\
    $GPTN_7$ & $\langle C_{dn}, I_{na}, Y \rangle^R$ & inconsistent network packet, deviation exists \\
    $GPTN_8$ & $\langle C_{dn}, I_{na}, N \rangle^R$ & inconsistent network packet, deviation does not exist \\\hline
    $LPTN_{1}$ & $\langle F/S, ED \rangle$ & expected deviation exists \\
    $LPTN_{2}$ & $\langle F, UED \rangle$ & unexpected deviation exists for float node \\
    $LPTN_{3}$ & $\langle S, UED \rangle$ & unexpected deviation exists for string node \\
    $LPTN_{4}$ & $\langle F/S, \top \rangle$ & log not available, or no deviation exists in the node \\
    $LPTN_{5}$ & $\langle Others, \top \rangle$ & function call or object (needed to maintain analytical flow) \\\hline
    \end{tabular}}
    \label{tab:new_patterns}
    \end{adjustwidth}
\end{table}

\section{Case Study}\label{sec:case_studies}
In this case study, we utilize a public DERMS application program repository\footnote{\url{https://github.com/GRIDAPPSD/gridappsd-soap-server/blob/master/soapServer/soap_server.py}} to evaluate the feasibility of our proposed techniques. In the codebase, there are several crucial virtual physical variables: \texttt{equipment\_id4dispatch}, \texttt{equipment\_name}, \texttt{equipment\_p}, \texttt{equipment\_q}, \texttt{equipment\_p\_meas}, \texttt{equipment\_q\_meas}, \texttt{equipment\_maxIFault}, \texttt{equipment\_ratedS}, \texttt{equipment\_ratedU}, and \texttt{equipment\_type}. These variables represent the unique ID, name, expected active power, expected reactive power, measured active power, measured reactive power, maximum fault current, rated apparent power (the maximum power capacity under normal conditions), rated apparent voltage, and the type of each of equipment, respectively. After analyzing the application program, a set of event types is generated (referred to Table~\ref{tab:events}). It should be noted that whereas the italic font ({\em{E}}) used in the table represents an event type, the non-italic font (E) used in the analysis represents an event instance. 

For each failure incident, a sequence of events can be extracted from in the corresponding log files. At the beginning of the analysis, the analyst can utilize this sequence to find the specific tainted path(s) in the dependency graph and ignore the unrelated ones. After that, the analyst goes through every node in that path and searches for matched patterns. Finally, a decision on the root cause is made based on the pattern-matching results. 

The following case studies show the practical use of this approach. The corresponding sequence of synthesized events is stored in CSV format and can be found in our dataset repository{\footnote{https://github.com/MdShamim097/Dataset-Events-Sequence.git}}. The CSV files are `events\_seq\_1', `events\_seq\_2', `events\_seq\_3', and `events\_seq\_4', representing the event sequences for the four case studies, aligned with their respective order. Each row in these CSV files represents an event instance, and the four columns denote the ID, type, recorded value, and description of the instance, respectively. For example, the row \texttt{2, E1, 1972301.79, `magnitude' sensing event} represents an instance of type E1, corresponding to a magnitude sensing event with the value $1972301.79$. We generated the synthesized events based on how such events are identified in the real world. It should be noted that events related to PMU measurements are collected from IEEE C37.118 direct capture files, network events are extracted from the relevant data fields of the PCAP (packet capture) file, and cyberspace events are collected from the corresponding DERMS application logs, which are usually saved as a TXT file.
\\ \\
\noindent{{\textbf{Case 1.}}} Consider a scenario in which equipment functions unexpectedly and significant discrepancies in their power measurements are monitored. Consequently, some equipment has been damaged, and an excessive efficiency loss is observed in the power system. In the ground truth, inconsistency is observed in the field measurements due to a system fault, and an ADS has detected something abnormal. 
\begin{table}[htbp]
\centering
\caption{{The event types used in our case studies}}
 \begin{adjustwidth}{-0.5cm}{}
\def\arraystretch{1}
    \resizebox{0.5311\textwidth}{!}{
\begin{tabular}{ll}
\hline\multicolumn{1}{c}{\textbf{\#}} & \multicolumn{1}{c}{\textbf{Description}} \\\hline
\em{E1} & a DER device or a virtual physical variable is found abnormal by an ADS \\
\em{E2} & response\_obj is consistent \\
\em{E3} & values in one or more fields of response\_obj are inconsistent \\
\em{E4} &  conditional check is passed \\
\em{E5} &  corresponding virtual physical variables are manipulated    \\
\em{E6} &  `equipment\_p\_meas' is consistent \\
\em{E7} &  `equipment\_q\_meas' is consistent \\
\em{E8} &  relevant intermediate mathematical calculations are inconsistent  \\
\em{E9} &  no manipulation in the mathematical calculations \\
\em{E10} & `forward\_value' is inconsistent \\
\em{E11} & `reverse\_value' is inconsistent \\
\em{E12} & active and reactive powers are consistent \\
\em{E13} & corresponding difference messages are altered \\
\em{E14} & corresponding difference messages are consistent \\
\em{E15} & `message' containing unexpected values is sent to the physical side \\ \hline
\end{tabular}}
\label{tab:events}
\end{adjustwidth}
\end{table}

\begin{figure*}[htbp]
    \begin{adjustwidth}{-0.711cm}{} 
        \includegraphics[scale=0.3311]{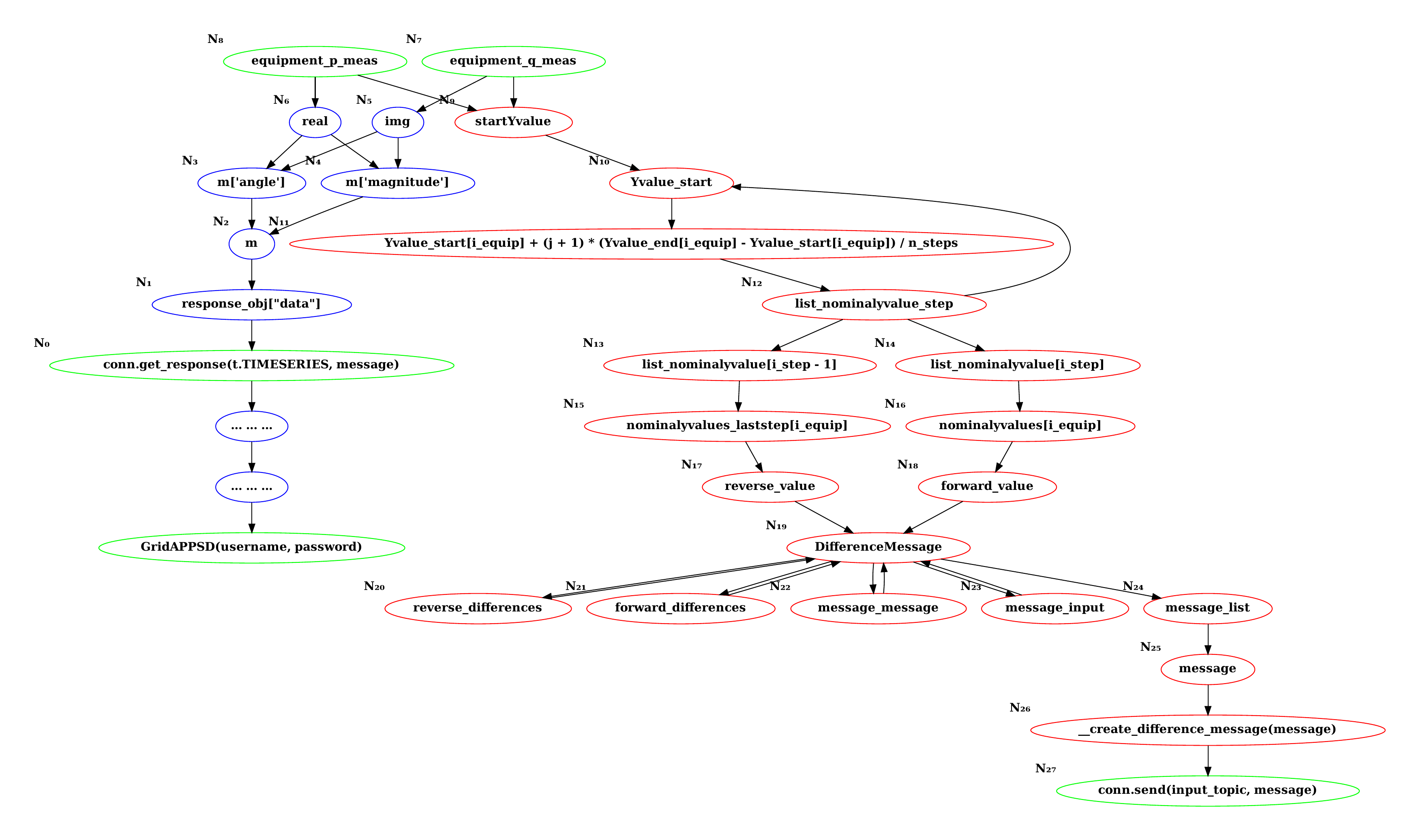}
    \end{adjustwidth}
    \vspace{-10pt}
    \caption{Dependency graph for case study.}
    \label{fig:cs_sim}
\end{figure*}

To investigate the cause of the system failure, a security analyst is assigned. After preliminary evaluation of the incident report, he finds that the ``angle" and ``magnitude" values of one equipment are inconsistent. Therefore, he looks into the codebase to get the appropriate program variable(s). Since only \texttt{equipment\_p\_meas} and \texttt{equipment\_q\_meas} are the virtual physical variables related to these measurements, the security specialist generates a dependency graph starting from these two variables according to the technique proposed in Section~\ref{sec:generation}. After initial generation of the upper subgraph (USG), two potential sources \texttt{GridAPPSD(username, password)} and \texttt{conn.get\_response(t.TIMESERIES, message)} are found using Heuristic at Section~\ref{sec:generation}. Then, the lower subgraph (LSG) is constructed using a forward tracking strategy that halts at the sink \texttt{conn.send(input\_topic, message)}. After that, the graph simplification algorithm is applied, which deducts $16$ nodes from the original graph ($57 \rightarrow 41$), achieving 28.07\% node reduction. The final graph is illustrated in Figure~\ref{fig:cs_sim}. For simplicity of further analysis, some intermediate nodes are skipped in the USG, and node numbering is started from the API request \texttt{conn.get\_response(t.TIMESERIES, message)}. Here, nodes in the USG and LSG are colored blue and red, respectively. In addition, the virtual physical variables, sources, and sinks are represented with yellow-colored nodes. \\
\begin{figure}[ht]
    \centering
    \subfloat[]{
        \includegraphics[scale=0.35]{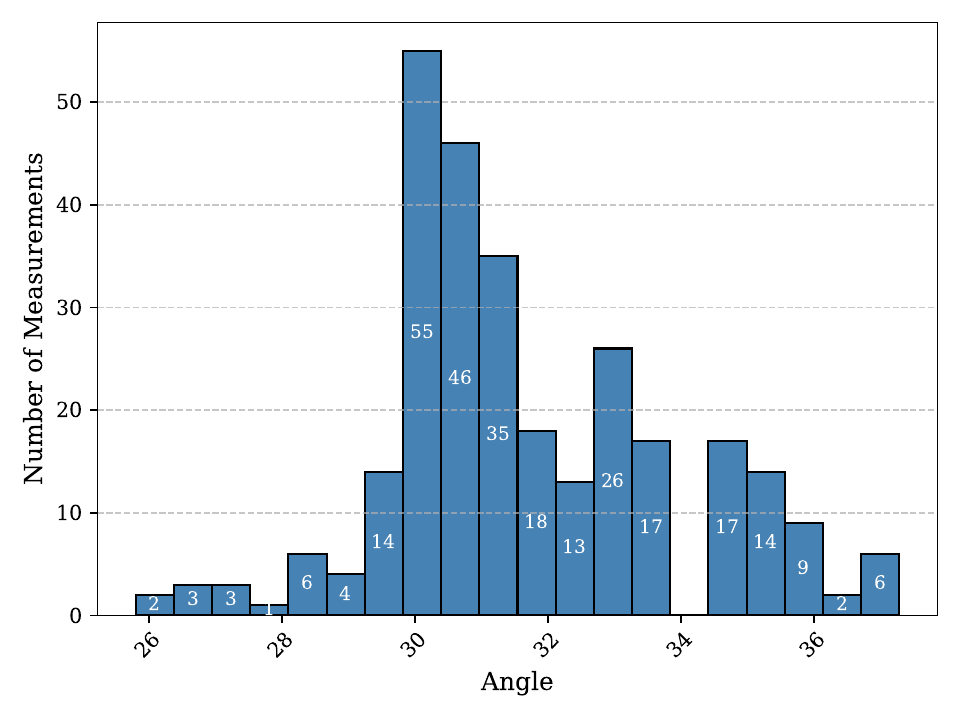}
        \label{fig:his_ang}
    }
    \hfill
    \subfloat[]{
        \includegraphics[scale=0.35]{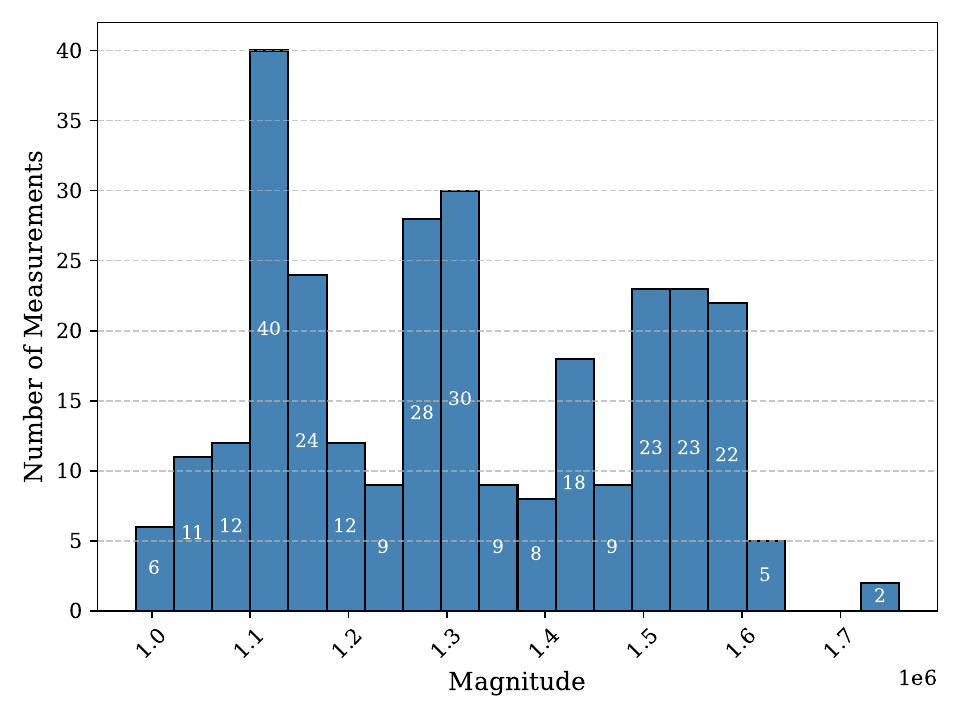}
        \label{fig:his_mag}
    }
    \caption{Histograms of (a) angle and (b) magnitude of the equipment}
\end{figure}
After pruning the dependency graph, the analyst uses the recorded sequence of events to identify the specific tainted path related to the incident. According to Table I, the analyst figures out that the subpath $N_{12} \rightarrow N_{13} \rightarrow N_{15} \rightarrow N_{17} \rightarrow N_{19}$ is not related to the incident. So, these nodes are not explored during further analysis. Next, each node along the related path is traversed to match a set of patterns based on the strategies described in Section~\ref{sec:patterns}.

From `events\_seq\_1' file of the dataset, ``angle" and ``magnitude" of the PMU measurement are $25.345$ and $1972301.79$, respectively (the sensing events). Besides, identical values are observed in the network packet as well as in the application log (network packet message event and response object, respectively). Based on this observation, the analyst generates two histograms to monitor whether these values are deviated from the `benign' ones or not. According to Figures~\ref{fig:his_ang} and~\ref{fig:his_mag}, the observed angle can be allocated to the leftmost bin; whereas, the magnitude cannot be assigned to any of them. Therefore, it is clear that \textit{deviation} exists in the PMU measurement, which matches with the global pattern $GPTN_5\equiv \langle C_{dn}, C_{na}, Y \rangle$ and the matched event instance is E1. In particular, the histograms are generated using a log collected from the National Renewable Energy Laboratory (NREL) that contains `benign' values of all essential measurements. After that, the analyst continues exploring the selected path and looks for potential matches of local patterns. The first node in the graph is a function call, whose corresponding pattern is $LPTN_5$. The next node is the response object, where the global pattern was found. Based on the observed values, the \textit{deviation} at this node can be considered as high (H). Since measurement information is carried from node $N_2$ to $N_9$, their executed values are arguably available in the application log. In addition, the observed deviations are as expected, that is $\Delta_X \rightarrow \Delta_Y$ since no program memory is corrupted. Thus, the matched patterns are $LPTN_1\equiv\langle F, ED \rangle$, and these patterns represent the occurrence of event instance E5. Notably, even if any log is not available, for example \texttt{startYvalue}, $\Delta_Y$ can be estimated because of their simple relations and code structures. However, from \texttt{Yvalue\_start} (node $N_{10}$) to \texttt{nominalyvalues[i\_equip]} (node $N_{16}$), logs are not available as they carry mathematical calculations. Although the analyst may try to estimate their values, it is not always possible due to their complex dependency. So, the deviations for these nodes are taken as unknown (pattern: $LPTN_4$). Importantly, such uncertainty of values does not affect the analysis even if any malicious activities take place within these nodes (e.g., memory corruption), because the contaminated values will ultimately propagate through the subsequent nodes to the end. After \texttt{nominalyvalues[i\_equip]}, next related node is \texttt{forward\_value}. Since it contains the measured active power value, its historical logs are certainly available, and the degree of deviation can be obtained using the histogram approach
In this case, the deviation found is $ED$ (pattern, $LPTN_1 \equiv\langle F, ED \rangle$) since no memory corruption has occurred. In addition, since the deviations exist at this node, event instance E10 is matched. The next node is an object (\texttt{DifferenceMessage}), which is matched with $LPTN_5$ according to the definition. Other nodes, especially from $N_{21}$ to $N_{25}$, are related to dispatching. So, their logs are always available, and the corresponding $\Delta_Y$ can be easily calculated. Consequently, the deviation shows $\Delta_X \rightarrow \Delta_Y$ characteristics, which indicate the event instances E13 and E15, and matches pattern $LPTN_1\equiv\langle S, ED \rangle$. In this way, E1$\rightarrow$E3$\rightarrow$E5$\rightarrow$ E10$\rightarrow$E13$\rightarrow$E15 is observed as the sequence of event instances for this failure incident.

After getting the pattern matching results, the analyst makes the final reasoning as follows: first, only one global pattern is found. Then, it is propagated through the program according to the deviation that is expected. In addition, no unexpected deviation is observed during the analysis. Therefore, the analyst can conclude that a system fault is the most likely root cause of the system failure. In this way, our proposed graph construction and pattern matching techniques facilitate the task of distinguishing system faults and cyberattacks in the CPS domain, especially in DER systems. The synthesized reasoning is shown in Table~\ref{tab:evaluation}.
\\ \\ 
\noindent{{\textbf{Case 2.}}} In this test case, an FDI attack is treated as the ground truth of system failure, and the corresponding recorded event sequence is shown in the `events\_seq\_2' file. Using this sequence, the analyst proceeds with the related tainted path. In this case, the ``angle" and ``magnitude" of the field measurements are found to be $30.7597$ and $1156739.528$, respectively. On the other hand, $30.7597$ and $1430307.59$ are found in the corresponding network packet and application log. From these values, it is clear that $v^d_x \ne v^n_x$ and $v^n_x = v^a_x$, which indicates the possibility of a suspicious act on the communication side. From the histograms illustrated in Figures~\ref{fig:his_ang} and~\ref{fig:his_mag}, the observed ``angle" can be allocated to one of the largest bins (not deviated much). However, the ``magnitude" is tracked to the twelfth bin, which can be considered as \textit{moderate deviation} (M). Thus, the analyst identifies $GPTN_1 \equiv \langle I_{dn} , C_{na}, Y\rangle$ as the global pattern and $LPTN_1 \equiv \langle F, ED \rangle$ as the local pattern (for node $N_1$). Regarding the analysis in the local region (application program), the matched patterns are similar to test case 1, as no direct interruption (e.g., memory corruption) takes place in the program region. However, the degree of deviations in the dependent nodes ($\Delta_X, \Delta_Y$) is different since this time $\Delta_X=M$ for the first node (response object), rather than $\Delta_X=H$. Although the observed sequence is similar to the previous test case, they represent different event instances and hence different incidents. Importantly, the matched global pattern plays a significant role in differentiating the root causes. All the matched patterns are shown in Table~\ref{tab:evaluation}.
\\ \\
\noindent{{\textbf{Case 3.}}} The ground truth for this test case is a memory corruption attack, where the attacker has manipulated one of the memory locations to change the reactive power measurements. According to the recorded sequence in the `events\_seq\_3' file, and the histograms in Figures~\ref{fig:his_ang} and~\ref{fig:his_mag}, \textit{no deviation} is found in these values, matching the global pattern $GPTN_4 \equiv \langle C_{dn}, C_{na}, N \rangle$. After initial selection based on the sequence of recorded events, the analyst suspects the subpath $N_{12} \rightarrow N_{13} \rightarrow N_{15} \rightarrow N_{17} \rightarrow N_{19}$ to be manipulated. So, he avoids analyzing the nodes $N_{12} \rightarrow N_{14} \rightarrow N_{16} \rightarrow N_{18} \rightarrow N_{19}$, and continues matching patterns in the selected path. At node $N_1$ (the response object), the pattern is $LPTN_1  \equiv \langle F, ED \rangle$. Importantly, here $\Delta=0$ since no deviation was observed during the global analysis, the event instance is E2. In the following nodes from $N_2$ to $N_9$, the same patterns $LPTN_1$ are found with $\Delta_X (=0) \rightarrow \Delta_Y(=0)$, which is expected and the found event instances are E6 and E7. However, at node $N_{17}$ containing {\texttt{reverse\_value}}, an unexpected deviation $(\Delta_X \rightarrow \Delta^{\prime}_Y)$ is observed (see 11$^{th}$ item in Table III), which indicates an occurrence of manipulation at this node or any of its predecessors on which this node has data dependency. Thus, the matched pattern is $LPTN_2 \equiv \langle F, UED \rangle$ and the corresponding matched event instance is E11. Since nodes $N_{19}-N_{20} \text{ and } N_{22}-N_{25}$ have data dependency with node $N_{17}$, their values are also unexpectedly changed. However, the matched patterns are $LPTN_3  \equiv \langle S, UED \rangle$ as these nodes contain string-typed values. Therefore, the observed event sequence in this case is: E1$\rightarrow$E2$\rightarrow$E6$\rightarrow$E7$\rightarrow$E11$\rightarrow$E13$\rightarrow$E15. Finally, reasoning can be done as follows: no deviation is found in the global region, but several unexpected deviations are observed in the local region (application program), which is a prominent indicator of a memory corruption attack. Therefore, the analyst can conclude a memory corruption is the root cause of system failure.
\\ \\ 
\noindent{{\textbf{Case 4.}}} In this test, the ground truth is also a memory corruption attack; but, the attack takes place in any of the intermediate nodes, where mathematical calculations are manipulated to alter the active power measurements. Specifically, the observed values in the field measurements, network packet, and application log are $36.173$ and $1243590.094$, respectively (see first four items in the `events\_seq\_4' file). Since these values remain consistent on both physical and cyber sides, the analyst further checks for potential deviations and their degrees. According to the histograms in Figures~\ref{fig:his_ang} and~\ref{fig:his_mag}, both values are located in two considerably moderate bins. Thus, \textit{deviation} (especially, M) may exist, which matches the global pattern $GPTN_5 \equiv \langle C_{dn}, C_{na}, Y \rangle$. According to the recorded events sequence in Table IV, the analyst selects the subpath $N_{12}\rightarrow N_{14}\rightarrow N_{16}\rightarrow N_{18}\rightarrow N_{19}$, and neglects the opposite one. In the local region, the patterns up to node $N_{16}$ are exactly identical to the previous test (case 3), except $\Delta_X$ and $\Delta_Y$ are not $0$ since deviation (expected or unexpected) is observed both in global and local portions. However, at node $N_{18}$, deviation is observed in the active power value while comparing the histogram of `benign" values. This indicates that memory corruption has occurred either at this node or any of the previous intermediate nodes. Consequently, the subsequent dependent nodes $N_{19}, N_{22}, N_{23}, N_{24}, \text{ and } N_{25}$ are also deviated. Regarding the event sequence, E1$\rightarrow$E3$\rightarrow$E5$\rightarrow$E10$\rightarrow$E13$\rightarrow$E15 is matched, which verifies the detection of related event instances. Interestingly, though the event sequence filters out the unrelated tainted path in the same way as in the previous case, they are substantially different and represent separate incidents. During the final reasoning, two major observations can be made: first, the matched global pattern indicates a slight deviation in response-related values; second, some expected and unexpected deviations are observed in the program region. Since alternation in the program side strongly justifies a memory corruption attack and the observed deviation in the global part does not necessarily indicate a system fault or FDI, the analyst can conclude memory corruption as the root cause of the system failure.
\begin{table*}[htbp]
\caption{Test case analysis}
\centering
\def\arraystretch{1}
    \resizebox{1\textwidth}{!}{
    \begin{threeparttable}
\begin{tabular}{cllcc}
\hline \textbf{\#}   & \multicolumn{1}{c}{\textbf{Pattern Matching Results}} & \multicolumn{1}{c}{\textbf{Nodes and Reasoning}}& \textbf{Matched Sequence}   & \textbf{Conclusion}   \\\hline
\multirow{7}{*}{1} & $GPTN_5$  & $N_0$: $LPTN_5$ (not applicable, function call) & \multirow{7}{*}{\begin{tabular}[c]{@{}c@{}}E1$^{s1}\rightarrow$E3$\rightarrow$E5$\rightarrow$\\ E10$\rightarrow$E13$\rightarrow$E15\end{tabular}} & \multirow{7}{*}{\begin{tabular}[c]{@{}c@{}}System Fault\end{tabular}} \\
& No $GPTN_1$ or $GPTN_2$ & $N_1$: $GPTN_5$ (response object itself; practically first node)  & &      \\
& No $GPTN_3$ or $GPTN_4$ & $N_2 - N_9$: $LPTN_{1}$ (response related) $(\Delta \ne 0)$  & &      \\
& No $GPTN_7$ or $GPTN_8$ & $N_{10} - N_{12}, N_{14}, N_{16}$: $LPTN_4$ (log not available; hard to estimate) & &     \\
& $LPTN_1$   & $N_{18}, N_{21} - N_{25}$: $LPTN_{1}$ (dispatch related) & &      \\
& No $LPTN_{2}$ or $LPTN_{3}$  & $ N_{19}$: $LPTN_5$ (not applicable; object name) & &  \\
& Some $LPTN_{4}$ and $LPTN_{5}$ (nothing abnormal)   & $ N_{26}, N_{27}$: $LPTN_5$ (not applicable; function call)  & &       \\\hline 

\multirow{7}{*}{2} & $GPTN_1$  & $N_0$: $LPTN_5$ & \multirow{7}{*}{\begin{tabular}[c]{@{}c@{}}E1$^{s2}\rightarrow$E3$\rightarrow$E5$\rightarrow$\\ E10$\rightarrow$E13$\rightarrow$E15\end{tabular}} & \multirow{7}{*}{\begin{tabular}[c]{@{}c@{}}FDI\end{tabular}} \\
& No $GPTN_3$ or $GPTN_4$ & $N_1$: $GPTN_1$   & &   \\
& No $GPTN_5$ or $GPTN_6$ & $N_2 - N_9$: $LPTN_{1}$ $(\Delta \ne 0)$ & &    \\
& No $GPTN_7$ or $GPTN_8$ & $N_{10} - N_{12}, N_{14}, N_{16}$: $LPTN_4$ & &     \\
& $LPTN_1$   & $N_{18}, N_{21} - N_{25}$: $LPTN_{1}$ & &   \\
& No $LPTN_{2}$ or $LPTN_{3}$  & $ N_{19}$: $LPTN_5$  & &\\
& Some $LPTN_{4}$ and $LPTN_{5}$    & $ N_{26}, N_{27}$: $LPTN_5$   & &   \\\hline 

\multirow{9}{*}{3} & $GPTN_4$  & $N_0$: $LPTN_5$  & \multirow{9}{*}{\begin{tabular}[c]{@{}c@{}}E1$\rightarrow$E2$\rightarrow$E6$\rightarrow$\\ E7$\rightarrow$E11$\rightarrow$E13$\rightarrow$E15\end{tabular}} & \multirow{9}{*}{\begin{tabular}[c]{@{}c@{}}Memory Corruption\end{tabular}} \\
& No $GPTN_1$ or $GPTN_2$ & $N_1$: $GPTN_4$  & &     \\
& No $GPTN_5$ or $GPTN_6$ & $N_2 - N_9$: $LPTN_{1} (\Delta=0)$    & &     \\
& No $GPTN_7$ or $GPTN_8$ & $N_{10} - N_{13}, N_{15}$: $LPTN_4$ & &    \\
& $LPTN_1$   & $N_{17}$: $LPTN_{2}$ (dispatch related; unexpected) & &      \\
& One $LPTN_{2}$  & $N_{20}, N_{22} - N_{25}$: $LPTN_{3}$ (dispatch related; unexpected) & &     \\
& Multiple $LPTN_{3}$ & $ N_{19}$: $LPTN_5$  & & \\
& Some $LPTN_{4}$ and $LPTN_{5}$ & $N_{21}$: $LPTN_1$ (dispatch related; expected) & &     \\
&  & $N_{26}, N_{27}$: $LPTN_5$ & &   \\\hline 

\multirow{9}{*}{4} & $GPTN_5$  & $N_0$: $LPTN_5$  & \multirow{9}{*}{\begin{tabular}[c]{@{}c@{}}E1$\rightarrow$E3$\rightarrow$E5$\rightarrow$\\ E10$\rightarrow$E13$\rightarrow$E15\end{tabular}} & \multirow{9}{*}{\begin{tabular}[c]{@{}c@{}}Memory Corruption\end{tabular}} \\
& No $GPTN_1$ or $GPTN_2$ & $N_1$: $GPTN_5$  & &   \\
& No $GPTN_3$ or $GPTN_4$ & $N_2 - N_9$: $LPTN_{1} (\Delta \ne 0)$    & &    \\
& No $GPTN_7$ or $GPTN_8$ & $N_{10} - N_{16}$: $LPTN_4$ & &     \\
& $LPTN_1$   & $N_{18}$: $LPTN_{2}$ & &    \\
& One $LPTN_{2}$  & $N_{21} - N_{25}$: $LPTN_{3}$ (dispatch related; unexpected) & &  \\
& Multiple $LPTN_{3}$ & $ N_{19}$: $LPTN_5$  & & \\
& Some $LPTN_{4}$ and $LPTN_{5}$ & $N_{20}$: $LPTN_1$ (dispatch related; expected) & &      \\
&  & $N_{26}, N_{27}$: $LPTN_5$ & &  \\\hline 

\multirow{7}{*}{5} & $GPTN_1$ (\it{false-positive})  & $N_0$: $LPTN_5$ & \multirow{7}{*}{\begin{tabular}[c]{@{}c@{}}E1$^{s2}\rightarrow$E3$\rightarrow$E5$\rightarrow$\\ E10$\rightarrow$E13$\rightarrow$E15\end{tabular}} & \multirow{7}{*}{\begin{tabular}[c]{@{}c@{}}FDI (\it{false-classification})\end{tabular}} \\
& No $GPTN_3$ or $GPTN_4$ & $N_1$: $GPTN_1$   & &   \\
& No $GPTN_5$ or $GPTN_6$ & $N_2 - N_9$: $LPTN_{1}$ $(\Delta \ne 0)$ & &    \\
& No $GPTN_7$ or $GPTN_8$ & $N_{10} - N_{12}, N_{14}, N_{16}$: $LPTN_4$ & &     \\
& $LPTN_1$   & $N_{18}, N_{21} - N_{25}$: $LPTN_{1}$ & &   \\
& No $LPTN_{2}$ or $LPTN_{3}$  & $ N_{19}$: $LPTN_5$  & &\\
& Some $LPTN_{4}$ and $LPTN_{5}$    & $ N_{26}, N_{27}$: $LPTN_5$   & &   \\\hline 
\end{tabular}
\end{threeparttable}}
\label{tab:evaluation}
\end{table*}
\\ \\ 
\noindent{{\textbf{Case 5: A False Classification Scenario.}}} To explore a possible false classification produced by our approach, we consider a scenario where the ground truth is a system fault. However, to mislead the pattern-based analysis, an attacker slightly modifies the packet field corresponding to ``magnitude", resulting in a false positive pattern being matched during the analysis while a true pattern remains undetected. Specifically, suppose that the ``angle" and ``magnitude" of the PMU measurements are $30.7597$ and $1972301.79$, respectively (the sensing events). In contrast, $30.7597$ and $1896739.424$ are found in the corresponding network packet and application log. Since $v^d_x \ne v^n_x$ and $v^n_x = v^a_x$, only $GPTN_1$ is matched from the global region. Although the ``magnitude" observed in the PMU measurements ($1972301.79$) deviates from the benign histogram logs, the global pattern $GPTN_5$ is not matched because the field, network, and application logs are no longer consistent. Consequently, $GPTN_1$ is matched while $GPTN_5$ remains undetected, indicating the possibility of suspicious activity on the communication side. Since no deviation is observed in the local region (program), the analyst concludes that the root cause is FDI. In the ideal case, the matched pattern would be $GPTN_5$, leading to a system fault diagnosis. However, because the attack prevents the system-fault-related global pattern from being matched and instead triggers the FDI-related pattern, the subsequent analyst reasoning -- performed after the pattern-matching stage -- results in an incorrect classification.
\\ \\
\noindent{\underline{Concluding Remarks.}} The conducted test cases bear clear testimony to the effectiveness of our proposed approach. Although all test cases reflect the significance of the co-existence of global and local patterns, each conveys a distinctive implication. More specifically, the first two cases emphasize the necessity of global patterns, especially how they effectively capture the distinct characteristics of individual causes. On the other hand, the importance of local patterns is demonstrated in the last two test cases. Notably, the fourth case demonstrates the necessity of this type of pattern to make the appropriate decision even when the matched global pattern indicates something else. In this way, the resulting differentiation framework becomes more rigorous and credible. However, it may be susceptible to false classification when an attacker intentionally manipulates the pattern-matching process, as demonstrated in the final case study.
\\ \\ 
\noindent{\underline{Comments on Benchmark Evaluation.}} Since existing works do not integrate physical side measurements and cyberspace information, no appropriate baseline is available to compare. Nevertheless, it should be noticed that the pattern matching mechanism can be scaled to a large-scale evaluation benchmark where a large number of recorded sequences of events are available. Using such a benchmark, performance metrics such as accuracy and F1 scores can be quantitatively measured. In our future work, we would collaborate with a DER industry partner and develop such a benchmark.

\section{Discussion}\label{sec:discussion}
In this work, we have introduced a semi-automatic approach to address the differentiation problem between undetected faults and cyberattacks in DER systems from a different perspective, specifically by jointly analyzing cyberspace and physical-side information. To evaluate the efficacy of the proposed approach, several case studies have been conducted, demonstrating its feasibility and effectiveness while mitigating key limitations of existing physical law-based and physical measurement-driven methods.
\\ \\ 
\noindent \textbf{Complexity Analysis of the Algorithms.}

Algorithm~\ref{alg:graph_construction}: Let $|P_{var}|$ denote the number of virtual physical variables, $V$ the number of graph nodes, and $E$ the number of graph edges. In Step 1, USG construction consists of backward traversal, $O(V+E)$, and taint-source identification, $O(V+E)$. Since function-call, callback, and inter-procedural relationships are represented as graph edges, taint propagation through functions does not introduce additional asymptotic cost beyond graph traversal. Therefore, USG construction requires $O(|P_{var}|(V+E))$ time. Similarly, LSG construction requires $O(|P_{var}|(V+E))$. Thus, the overall complexity of Algorithm 1 is $O(|P_{var}|(V+E))+O(|P_{var}|(V+E))=O(|P_{var}|(V+E))$.

Algorithm~\ref{alg:usg_simplification}: Let $P$ denote the number of paths from $P_{var}$ to $T_{src}$. Line 3 requires $O(V)$. During Scan 1, the importance-update phase processes every path between virtual physical variables and taint sources. Since each path contains at most $V$ nodes, the worst-case complexity is $O(PV)$. Scans 2 and 3 each require $O(V+E)$. Therefore, the overall complexity of Algorithm 2 is $O(PV + V + E)=O(PV + E)$.

Algorithm~\ref{alg:lsg_simplification}: Computing incoming-edge counts requires $O(V+E)$. Similar to Algorithm 2, the importance-update phase in Scan 1 requires $O(PV)$ in the worst case. Scans 2 and 3 each require $O(V+E)$. Hence, the overall complexity of Algorithm 3 is $O(V+E)+O(PV)+O(V+E)+O(V+E)=O(PV + V + E)=O(PV + E)$.

Algorithm~\ref{alg:pattern_matching}: Let $P=|P_{rel}|$ denote the total number of incident-related paths, $V_x$ the number of response/dispatch-related variables, and $N_p$ the total number of node occurrences across $P_{rel}$. During global pattern matching, each iteration performs constant-time comparisons and lookups, resulting in $O(V_x)$. During local pattern matching, each operation, including histogram lookup, expected-value lookup, and node-type lookup, requires $O(1)$. Therefore, local matching requires $O(N_p)=\sum_{p\in P_{rel}} |p|$, yielding an overall complexity of $O(V_x+N_p)$.

Overall Complexity: Combining the complexities of graph construction, USG simplification, LSG simplification, and pattern matching yields $O(|P_{var}|(V+E)) + O(PV+E) + O(PV+E) + O(V_x+N_p) = O(|P_{var}|(V+E)+PV+E+V_x+N_p)$. Since $N_p=\sum_{p\in P_{rel}} |p| \le PV$ and $V_x \le V$, the pattern-matching complexity is asymptotically dominated by the graph-simplification phase. Therefore, the overall complexity can be simplified to $O(|P_{var}|(V+E)+PV)$, where $|P_{var}|$ is the number of virtual physical variables, $V$ and $E$ are the numbers of graph nodes and edges, respectively, and $P$ is the number of paths processed during graph simplification.
\\ \\
\noindent \textbf{Applicability Conditions.} Since the proposed differentiation framework is algorithmic, its applicability is primarily determined by the availability of the inputs required by the algorithms. In particular, the framework assumes the availability of (i) an incident report, (ii) a recorded event sequence together with PMU measurements, network packet (PCAP) files, and sufficient DERMS application logs generated during program execution, and (iii) historical benign logs. When one or more of these inputs are unavailable, incomplete, or of insufficient quality, the effectiveness of the framework may be affected, as discussed below.
\\ \\ 
\noindent \textbf{Impact of Measurement Inaccuracy.} Although our approach exhibits effectiveness for the root cause analysis, in practical DER systems, issues such as measurement noise, log loss, and time-synchronization errors among different systems are not uncommon. To analyze these inaccuracies, first, we recall that the proposed pattern-matching procedure relies on histograms constructed from benign logs. Measurement noise may reduce the accuracy of this procedure, as deviations in the observed values can shift the histogram distributions and lead to incorrect pattern matches. In contrast, log loss may result in some important patterns being missed. The impact of such missing patterns depends on whether the loss is occurred in the global region or local region (i.e, the application program). In the former case, the root-cause analysis may be affected. In the latter case, the impact is often recoverable, since the local pattern $LPTN_4$ and the notions of expected and unexpected deviations introduced in Section~\ref{sec:patterns} are specifically designed to handle such situations. Regarding time-synchronization errors, the proposed patterns are defined using the relative ordering of timestamps (e.g., $t_x^d < t_x^n < t_x^a$) rather than their absolute values. Consequently, moderate synchronization errors are unlikely to result in incorrect pattern matches. Nevertheless, some patterns may remain unmatched under severe synchronization errors, in which case the impact is similar to that of log-loss scenarios.

\section{Conclusion and Future Work}\label{sec:conclusion}
This paper introduces the first combined analysis of physical measurements and cyberspace information to differentiate faults and cyberattacks in DER systems. In particular, a non-trivial taint analysis technique is proposed to efficiently capture the propagation of untrusted data in DER application programs, along with devising a set of patterns to bridge the semantic gap between cyber and physical sides. During analysis, the generated graph is traversed to match a set of patterns, and the decision on the root cause of system failure is made based on the pattern-matching results. 

However, the proposed analysis framework still requires human intervention during the Incident-Related Path Selection and Root-Cause Reasoning stages of Phase II. Specifically, security analysts must leverage event sequences and his expertise to identify relevant paths and interpret pattern-matching results. One future direction could focus on further automating these stages, particularly by developing techniques that can automatically reason over pattern-matching results and infer the most likely root causes, thereby improving the scalability and efficiency of the overall analysis process.

\section*{Acknowledgment}
This work was partially supported by the U.S. Department of Energy under Grant DOE CESER RC-40125b-2023.

\begingroup\scriptsize
\makeatletter
\renewcommand\@openbib@code{\itemsep\z@}
\makeatother
\bstctlcite{IEEEexample:BSTcontrol}        
\bibliographystyle{IEEEtranS} 
\bibliography{doe}

@article{inc2_liang20162015,
  title={The 2015 Ukraine blackout: Implications for false data injection attacks},
  author={Liang, Gaoqi and Weller, Steven R and Zhao, Junhua and Luo, Fengji and Dong, Zhao Yang},
  journal={IEEE transactions on power systems},
  volume={32},
  number={4},
  pages={3317--3318},
  year={2016},
  publisher={IEEE}
}

@MISC{inc3_irgc2020,
    author = "{America's Cyber Defense Agency}",
    date = {2024},
    title = "{IRGC-Affiliated Cyber Actors Exploit PLCs in Multiple Sectors, Including US Water and Wastewater Systems Facilities}",
    HOWPUBLISHED = "https://www.cisa.gov/news-events/cybersecurity-advisories/aa23-335a",
    note = "Accessed: 13 January, 2025"
}

@article{gupta2022distinguishing,
  title={Distinguishing between cyber attacks and faults in power electronic systems—A noninvasive approach},
  author={Gupta, Kirti and Sahoo, Subham and Mohanty, Rabindra and Panigrahi, Bijaya Ketan and Blaabjerg, Frede},
  journal={IEEE Journal of Emerging and Selected Topics in Power Electronics},
  volume={11},
  number={2},
  pages={1578--1588},
  year={2022},
  publisher={IEEE}
}

@article{ao2016adaptive,
  title={Adaptive cyber-physical system attack detection and reconstruction with application to power systems},
  author={Ao, Wei and Song, Yongdong and Wen, Changyun},
  journal={IET Control Theory \& Applications},
  volume={10},
  number={12},
  pages={1458--1468},
  year={2016},
  publisher={Wiley Online Library}
}

@article{beg2021cyber,
  title={Cyber-physical anomaly detection in microgrids using time-frequency logic formalism},
  author={Beg, Omar Ali and Nguyen, Luan Viet and Johnson, Taylor T and Davoudi, Ali},
  journal={IEEE Access},
  volume={9},
  pages={20012--20021},
  year={2021},
  publisher={IEEE}
}

@article{sahoo2020resilient,
  title={Resilient synchronization strategy for AC microgrids under cyber attacks},
  author={Sahoo, Subham and Yang, Yongheng and Blaabjerg, Frede},
  journal={IEEE Transactions on Power Electronics},
  volume={36},
  number={1},
  pages={73--77},
  year={2020},
  publisher={IEEE}
}

@article{amini2017hierarchical,
  title={Hierarchical location identification of destabilizing faults and attacks in power systems: A frequency-domain approach},
  author={Amini, Sajjad and Pasqualetti, Fabio and Abbaszadeh, Masoud and Mohsenian-Rad, Hamed},
  journal={IEEE Transactions on Smart Grid},
  volume={10},
  number={2},
  pages={2036--2045},
  year={2017},
  publisher={IEEE}
}

@article{shen2024detection,
  title={Detection, differentiation and localization of replay attack and false data injection attack based on random matrix},
  author={Shen, Yuehao and Qin, Zhijun},
  journal={Scientific Reports},
  volume={14},
  number={1},
  pages={2758},
  year={2024},
  publisher={Nature Publishing Group UK London}
}

@article{ganjkhani2021integrated,
  title={Integrated cyber and physical anomaly location and classification in power distribution systems},
  author={Ganjkhani, Mehdi and Gilanifar, Mostafa and Giraldo, Jairo and Parvania, Masood},
  journal={IEEE Transactions on Industrial Informatics},
  volume={17},
  number={10},
  pages={7040--7049},
  year={2021},
  publisher={IEEE}
}

@article{sakhnini2021physical,
  title={Physical layer attack identification and localization in cyber--physical grid: An ensemble deep learning based approach},
  author={Sakhnini, Jacob and Karimipour, Hadis and Dehghantanha, Ali and Parizi, Reza M},
  journal={Physical Communication},
  volume={47},
  pages={101394},
  year={2021},
  publisher={Elsevier}
}

@article{madabhushi2023survey,
  title={A survey of anomaly detection methods for power grids},
  author={Madabhushi, Srinidhi and Dewri, Rinku},
  journal={International Journal of Information Security},
  volume={22},
  number={6},
  pages={1799--1832},
  year={2023},
  publisher={Springer}
}

@article{mo2011cyber,
  title={Cyber--physical security of a smart grid infrastructure},
  author={Mo, Yilin and Kim, Tiffany Hyun-Jin and Brancik, Kenneth and Dickinson, Dona and Lee, Heejo and Perrig, Adrian and Sinopoli, Bruno},
  journal={Proceedings of the IEEE},
  volume={100},
  number={1},
  pages={195--209},
  year={2011},
  publisher={IEEE}
}

@article{t4_arzt2014flowdroid,
  title={Flowdroid: Precise context, flow, field, object-sensitive and lifecycle-aware taint analysis for android apps},
  author={Arzt, Steven and Rasthofer, Siegfried and Fritz, Christian and Bodden, Eric and Bartel, Alexandre and Klein, Jacques and Le Traon, Yves and Octeau, Damien and McDaniel, Patrick},
  journal={ACM sigplan notices},
  volume={49},
  number={6},
  pages={259--269},
  year={2014},
  publisher={ACM New York, NY, USA}
}

@inproceedings{t5_sridharan2011f4f,
  title={F4F: taint analysis of framework-based web applications},
  author={Sridharan, Manu and Artzi, Shay and Pistoia, Marco and Guarnieri, Salvatore and Tripp, Omer and Berg, Ryan},
  booktitle={Proceedings of the 2011 ACM international conference on Object oriented programming systems languages and applications},
  pages={1053--1068},
  year={2011}
}

@article{tertytchny2020classifying,
  title={Classifying network abnormalities into faults and attacks in IoT-based cyber physical systems using machine learning},
  author={Tertytchny, Georgios and Nicolaou, Nicolas and Michael, Maria K},
  journal={Microprocessors and Microsystems},
  volume={77},
  pages={103121},
  year={2020},
  publisher={Elsevier}
}

@article{syfert2022integrated,
  title={Integrated approach to diagnostics of failures and cyber-attacks in industrial control systems},
  author={Syfert, Micha{\l} and Ordys, Andrzej and Ko{\'s}cielny, Jan Maciej and Wnuk, Pawe{\l} and Mo{\.z}aryn, Jakub and Kukie{\l}ka, Krzysztof},
  journal={Energies},
  volume={15},
  number={17},
  pages={6212},
  year={2022},
  publisher={MDPI}
}

@article{alam2015comprehensive,
  title={A comprehensive review of catastrophic faults in PV arrays: types, detection, and mitigation techniques},
  author={Alam, Mohammed Khorshed and Khan, Faisal and Johnson, Jay and Flicker, Jack},
  journal={IEEE Journal of Photovoltaics},
  volume={5},
  number={3},
  pages={982--997},
  year={2015},
  publisher={IEEE}
}

@inproceedings{zhao2011fault,
  title={Fault analysis in solar PV arrays under: Low irradiance conditions and reverse connections},
  author={Zhao, Ye and Lehman, Brad and de Palma, Jean-Fran{\c{c}}ois and Mosesian, Jerry and Lyons, Robert},
  booktitle={2011 37th IEEE Photovoltaic Specialists Conference},
  pages={002000--002005},
  year={2011},
  organization={IEEE}
}

@inproceedings{benninger2020anomaly,
  title={Anomaly detection by comparing photovoltaic systems with machine learning methods},
  author={Benninger, Moritz and Hofmann, Martina and Liebschner, Marcus},
  booktitle={NEIS 2020; Conference on Sustainable Energy Supply and Energy Storage Systems},
  pages={1--6},
  year={2020},
  organization={VDE}
}

@article{wang2022online,
  title={Online automatic anomaly detection for photovoltaic systems using thermography imaging and low rank matrix decomposition},
  author={Wang, Qian and Paynabar, Kamran and Pacella, Massimo},
  journal={Journal of quality technology},
  volume={54},
  number={5},
  pages={503--516},
  year={2022},
  publisher={Taylor \& Francis}
}

@inproceedings{tsai2020anomaly,
  title={Anomaly detection mechanism for solar generation using semi-supervision learning model},
  author={Tsai, Chia-Wei and Yang, Chun-Wei and Hsu, Feng-Ling and Tang, Hsih-Min and Fan, Nien-Chin and Lin, Cheng-Yang},
  booktitle={2020 Indo--Taiwan 2nd International Conference on Computing, Analytics and Networks (Indo-Taiwan ICAN)},
  pages={9--13},
  year={2020},
  organization={IEEE}
}
\endgroup
\appendix

\begin{appendices}

\subsection*{A. Taint Propagation Heuristic}
Heuristic~\ref{alg:heuristics_2} describes the additional taint propagation rules introduced in this work to extend classic taint analysis for handling function-call relationships. These rules govern propagation through function arguments, object references, inter-procedural flows, and callback functions.
\begin{algorithm}[htbp]
\footnotesize
\heuristiccaption{Taint Propagation through Function Calls}
\label{alg:heuristics_2}
\begin{algorithmic}[1]

\Require Upper Subgraph $USG$
\Ensure Updated taint propagation paths

\Statex {\color{lightgray}// \textit{Rule I: Backtracking through function calls}}
\ForAll{function calls}
    \State Propagate taint from each \emph{argument}.
    \If{an \emph{argument} is related to an \emph{object}}
        \State Propagate taint through the object.
    \EndIf
    \If{the function is bound to an \emph{object}}
        \State Treat the object as the receiver of taint.
    \EndIf
    \If{the flow is \emph{inter-procedural}}
        \State Start backtracking from the \texttt{return} statement.
    \EndIf
\EndFor

\Statex {\color{lightgray}// \textit{Rule II: Forward tracking through function calls}}
\ForAll{forward propagation steps}
    \If{the variable is an argument of an API call, command-line argument function,
    query function, or input function}
        \State Do not propagate through that argument variable.
        \State Continue propagation from the variable that assigns its value.
    \EndIf
    \If{the call crosses functions}
        \State Propagate through tainted parameters inside the \emph{callee} function.
    \EndIf
\EndFor

\Statex {\color{lightgray}// \textit{Rule III: Handling Callback Functions}}
\ForAll{\emph{callback} functions}
    \State Treat the callback as a special case since it is not directly called.
    \State Propagate from already tainted variables of the related object.
    \State Identify callback functions during forward tracking.
\EndFor

\end{algorithmic}
\end{algorithm}

\begin{figure}[htbp]
    \centering 
        \includegraphics[scale=0.5]{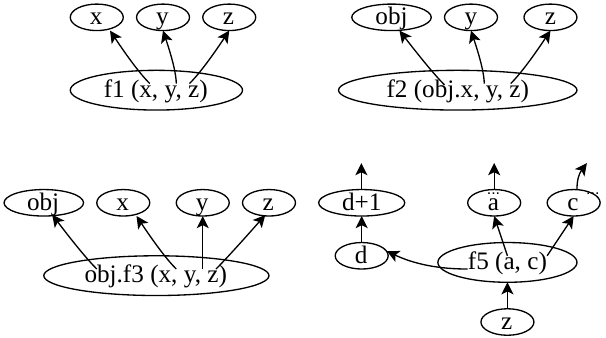}
    \caption{Handling propagation through function during backtracking}
    \label{fig:gen_example}
\end{figure}

Figure~\ref{fig:gen_example} illustrates several backtracking cases for propagation of function-calls. In the first case, the arguments {\tt x}, {\tt y}, and {\tt z} are propagated through {\tt f1(x,y,z)}. In the second case, the object {\tt obj} is also considered because it is passed as an argument in {\tt f2(obj,x,y,z)}. In the third case, the function is bound to an object, as in {\tt obj.f3(x,y,z)}, so the object is included in the propagation. The last case shows that, for {\tt f5(a,c)}, the following follows begins from the {\tt return} value {\tt z} and follows the dependency through {\tt d}, {\tt d+1}, and the input arguments.

Figure~\ref{fig:inter_flow} shows the corresponding inter-procedural code example. The function {\tt f5(x,y)} computes {\tt d=x + y}, updates it to {\tt d=d + 1} and returns {\tt d}. In {\tt f4(a,b,c)}, the returned value of {\tt f5(a,c)} is assigned to {\tt z}. Therefore, the backtracking from {\tt z} crosses the boundary of the function to {\tt f5} and continues through the returned variable {\tt d}.

Figure~\ref{fig:ind_conn} illustrates a code example for an indirect connection. Inside function {\tt f6(x,y,z)}, {\tt f7(y,z)} is called, while function {\tt f7} contains the virtual physical variable region. This creates an indirect relation between {\tt f6} and the virtual physical variable region through {\tt f7}. Such indirect connections are excluded from the proposed approach to avoid unnecessary complexity.
\begin{figure}[H] 
    \centering
    \begin{minipage}[t][4.5cm][t]{0.49\columnwidth} 
        \begin{lstlisting}[language=Python, label={lst:ex1}, basicstyle=\small\ttfamily]
def f5(x, y):
    d = x + y
    d = d + 1
    return d

def f4(a, b, c):
    ...
    z = f5(a, c)
    ...
        \end{lstlisting}
        \caption{An Inter-procedural Flow.}
        \label{fig:inter_flow}
    \end{minipage}
    \begin{minipage}[t][4.5cm][t]{0.49\columnwidth} 
        \begin{lstlisting}[language=Python, label={lst:ex2}, basicstyle=\small\ttfamily]
def f6(x, y, z):
    f7(y, z)
    ...

def f7(a, b):
    '''
    virtual physical variables region
    '''
        \end{lstlisting}
        \caption{An Indirect Connection.}
        \label{fig:ind_conn}
    \end{minipage}
\end{figure}

\subsection*{B. Taint Source Identification Heuristic}
Heuristic~\ref{alg:heuristics_1} describes the strategy used to identify taint sources within the upper subgraph. At the beginning of the heuristic, two observations are made: i) a user-defined function can never be a taint source, and ii) taint sources typically correspond to API calls, command-line argument handling functions, query processing functions, and various types of read functions. During backtracking, each node is assigned a weight according to its relationship with virtual physical variables and potential taint sources. Subsequently, forward tracking is performed along each path starting from a node with no outgoing edges, while nodes are progressively removed until a node satisfying two conditions is encountered: i) weight is at least one, and ii) exhibits characteristics of a possible taint source. Whenever such a node is found, the process is stopped.
\begin{algorithm}[htbp]
\footnotesize
\heuristiccaption{Identification of Taint Sources}
\label{alg:heuristics_1}
\begin{algorithmic}[1]

\Require Upper Subgraph ($USG$)
\Ensure Set of identified taint sources ($T_{src}$)

\Statex  {\color{lightgray}// \textit{Step 1: Node Weighting}}
\ForAll{nodes $N_k \in USG$ during backtracking}
    \If{$N_k \in P_{var}$}
        \State $W_k \gets 1$
    \ElsIf{$N_k \in T_{src}$}
        \State $W_k \gets W_{k+1} + 1$
    \Else
        \State $W_k \gets W_{k+1} + \frac{1}{2}$
    \EndIf
\EndFor

\Statex {\color{lightgray} // \textit{Step 2: Node Removal}}
\ForAll{forward-tracking paths in $USG$}
    \State Select a node with no outgoing edges.
    
    \While{nodes remain on the path}
        \State Remove the current node.
        
        \If{$W_k \geq 1$ \textbf{and}
            the node contains a feature of a possible taint source}
            \State Add the node to $T_{src}$.
            \State \textbf{break}
        \EndIf
    \EndWhile
\EndFor

\State \Return $T_{src}$

\end{algorithmic}
\end{algorithm}

\subsection*{C. USG Importance Update Heuristic}
The importance-update strategy used during USG simplification is described in Heuristic~\ref{alg:heuristics_3}. In particular, the heuristic assigns and updates node importance values based on their contextual roles in dependency propagation. Specifically, it prioritizes virtual physical variables, taint sources, assignment variables, expressions, array and object references, and user-defined functions, while assigning lower importance to auxiliary and intermediate nodes. 

\begin{figure}[htbp]
    \centering 
        \includegraphics[scale=0.5]{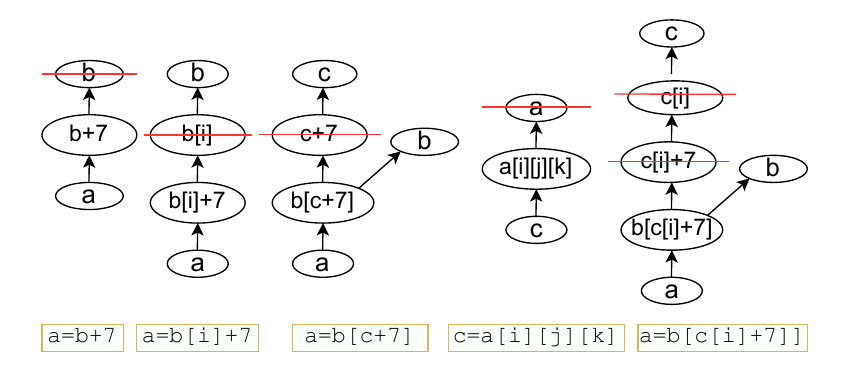}
    \caption{Illustrations of importance update and node removal rules for USG}
    \label{fig:usg_upd_example}
\end{figure}

Examples illustrating the importance-update rules are shown in Figure~\ref{fig:usg_upd_example}. From left to right, the first example corresponds to the statement {\tt{a=b+7}}, where the associated nodes are $a$, $b+7$, and $b$. Since $a$ and $b$ are auxiliary variables, the rule (Rule II) updates their importance by $\Delta imv_k(a)=1$, $\Delta imv_{k+1}(b+7)=1$, and $\Delta imv_{k+2}(b)=\alpha$. The second example illustrates the statement {\tt{a=b[i]+7}}, whose corresponding nodes are $a$, $b[i]+7$, $b[i]$, and $b$. According to the rule (Rule II), their importance values are increased by $1$, $1$, $\alpha$, and $1$, respectively. The third example considers {\tt{a=b[c+7]}}, where the associated nodes are $a$, $b[c+7]$, $b$, $c+7$, and $c$. Their importance values are updated to $imv_k(a)+=1$, $imv_{k+1}(b[c+7])+=1$, $imv_{k+2}(b)+=1$, $imv_{k+3}(c+7)+=\alpha$, and $imv_{k+4}(c)+=1$, based on Rule II. The fourth example shows the statement {\tt{a[i][j][k]}}, where the associated nodes are $a$ and $a[i][j][k]$. According to the rule (Rule III), $a[i][j][k]$ is prioritized over $a$, i.e., $imv_k(a[i][j][k])+=1$ and $imv_{k+1}(a)+=\alpha$. Finally, the fifth example illustrates the statement {\tt{a=b[c[i]+7]}}, where the associated nodes are $a$, $b[c[i]+7]$, $b$, $c[i]+7$, $c[i]$, and $c$. According to the rule (Rule IV), the intermediate nodes $c[i]+7$ and $c[i]$ are assigned lower importance by increasing their values by $\alpha$.
\begin{algorithm}[htbp]
\footnotesize
\heuristiccaption{USG Importance Update}
\label{alg:heuristics_3}

\begin{algorithmic}[1]

\Require Upper Subgraph $USG$
\Ensure Updated importance values $imv_k$

\Statex {\color{lightgray}// \textit{Rule I: Basic Rule}}
\ForAll{nodes $N_k \in USG$ during backtracking}
    \If{$N_k \in P_{var}$ or $N_k \in T_{src}$}
        \State $imv(N_k) \gets 1$
    \ElsIf{$N_k$ is an auxiliary variable ($V_s$) encountered for the first time}
        \State $imv(N_k) \gets 1$
    \EndIf

    \If{multiple consecutive nodes represent the same entity}
        \State $imv(\textit{earliest occurrence}) \pluseq 1$
        \State $imv(\textit{each subsequent occurrence}) \pluseq \alpha$
    \EndIf
\EndFor

\Statex {\color{lightgray}// \textit{Rule II: Expression Rule}}
\ForAll{expression-related nodes}
    \If{the expression consists only of auxiliary-variable operands}
        \State $imv(\textit{assignment variable}) \pluseq 1$
        \State $imv(\textit{expression node}) \pluseq 1$
        \State $imv(\textit{each operand node}) \pluseq \alpha$
    \ElsIf{the expression contains array or object references}
        \State $imv(\textit{assignment variable}) \pluseq 1$
        \State $imv(\textit{expression node}) \pluseq 1$
        \State $imv(\textit{reference expression}) \pluseq \alpha$
        \State $imv(\textit{base array/object}) \pluseq 1$
    \ElsIf{the expression is used within an array or object reference}
        \State $imv(\textit{assignment variable}) \pluseq 1$
        \State $imv(\textit{outer reference expression}) \pluseq 1$
        \State $imv(\textit{base array/object}) \pluseq 1$
        \State $imv(\textit{index variable}) \pluseq 1$
        \State $imv(\textit{inner expression node}) \pluseq \alpha$
    \EndIf
\EndFor

\Statex {\color{lightgray}// \textit{Rule III: Array/Object Referencing}}
\ForAll{array or object references}
    \State $imv(\textit{reference expression}) \pluseq 1$
    \State $imv(\textit{base array/object}) \pluseq \alpha$
\EndFor

\Statex {\color{lightgray}// \textit{Rule IV: Multiple Inner Referencing}}
\ForAll{nested array or object references}
    \State $imv(\textit{outer reference expression}) \pluseq 1$
    \State $imv(\textit{base array/object}) \pluseq 1$
    \State $imv(\textit{base object of each inner reference}) \pluseq 1$
    \State $imv(\textit{each intermediate node}) \pluseq \alpha$
\EndFor

\Statex {\color{lightgray}// \textit{Rule V: Function Calls}}
\ForAll{function-call nodes}
    \If{the function is built-in}
        \State $imv(\textit{built-in function node}) \pluseq \alpha$
        \State Preserve the assigned importance value during subsequent updates.
    \Else
        \State $imv(\textit{user-defined function/method}) \pluseq 1$
    \EndIf
\EndFor

\end{algorithmic}
\end{algorithm}

\subsection*{D. LSG Importance Update Heuristic}
Heuristic~\ref{alg:heuristics_4} describes the importance-update procedure used during the LSG simplification. Specifically, it prioritizes user-defined functions, the latest occurrence of consecutive nodes representing the same entity, and array or object references, while assigning lower importance to built-in functions and intermediate referencing nodes. 

\begin{figure}[htbp]
    \centering 
        \includegraphics[scale=0.61]{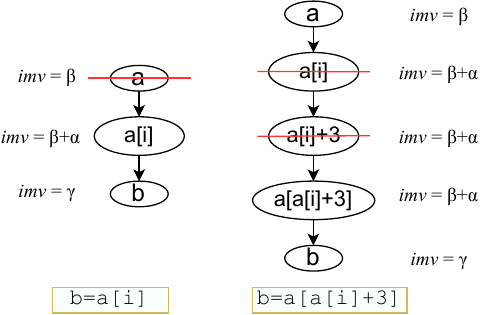}
    \caption{Illustration of importance update and node removal rules for LSG}
    \label{fig:lsg_upd_example}
\end{figure}

Examples related to the importance update are shown in Figure~\ref{fig:lsg_upd_example}. For instance, consider the statement {{\texttt{b=a[a[i]+3]}}}; where, the associated nodes are: $a$, $a[i]$, $a[i]+3$, $a[a[i]+3]$, and $b$. Let $imv_k(a)=\beta \text{ and } imv_{k+4}(b)=\gamma$, where $\beta, \gamma \in \mathbb{Z^+}$ since the importance value of a node is linearly proportional to the number of incoming edges and belongs to the set of positive integers: $\mathbb{Z^+}=\{1,2,3, ...\infty\}$. In accordance with the importance updates rule (Heuristic~\ref{alg:heuristics_4}), other values will be $imv_{k+1}(a[i])=imv_{k+2}(a[i]+3)=imv_{k+3}(a[a[i]+3])=\beta + \alpha$; thus identical. Therefore, nodes $a[i]$ and $a[i]+3$ will be removed from the graph, prioritizing nodes $a$ and $a[a[i]+3]$.
\begin{algorithm}[htbp]
\footnotesize
\heuristiccaption{LSG Importance Update}
\label{alg:heuristics_4}
\begin{algorithmic}[1]

\Require Lower Subgraph $LSG$, importance factor $\alpha$
\Ensure Updated importance values $imv_k$

\Statex {\color{lightgray}// \textit{Rule I: Function Calls}}
\ForAll{function-call nodes}
    \If{the function is built-in}
        \State $imv(\textit{built-in function}) \pluseq \alpha$
        \State Preserve the assigned importance value during subsequent updates.
    \Else
        \State $imv(\textit{user-defined function/method}) \pluseq 1$
    \EndIf
\EndFor

\Statex {\color{lightgray}// \textit{Rule II: Consecutive and Referencing Nodes}}
\ForAll{consecutive nodes representing the same entity}
    \State $imv(\textit{latest occurrence}) \pluseq 1$
    \State $imv(\textit{each previous occurrence}) \pluseq \alpha$
\EndFor

\ForAll{consecutive nodes containing array or object references}
    \State $imv(\textit{first reference}) \pluseq \alpha$
    \State $imv(\textit{subsequent references}) \gets imv(\textit{first reference})$
\EndFor

\end{algorithmic}
\end{algorithm}

\end{appendices}

\vskip -2\baselineskip plus -1fil

\begin{IEEEbiography}
[{\includegraphics[width=1in,height=1.25in,clip,keepaspectratio]{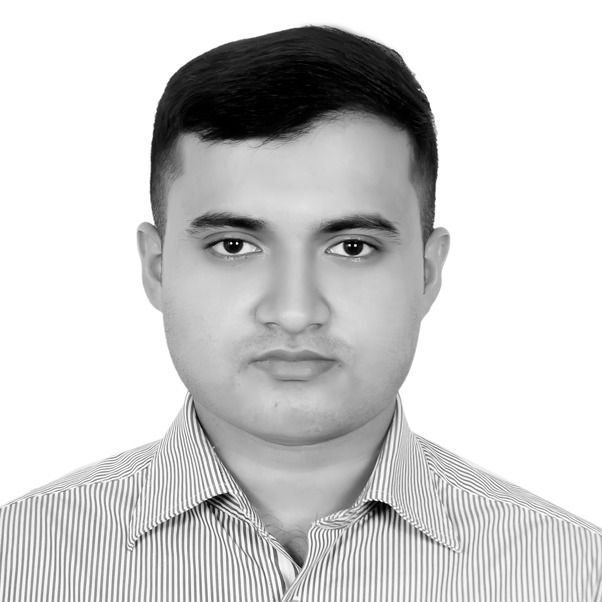}}]{Mohammad Shamim Ahsan}
is a Ph.D. student in the College of Information Sciences and Technology at The Pennsylvania State University. He received the B.Sc. degree in Computer Science and Engineering from the Bangladesh University of Engineering and Technology (BUET), Dhaka, Bangladesh, in 2023. From 2023 to 2024, he was a Lecturer with the Department of Computer Science and Engineering, United International University (UIU), Dhaka, Bangladesh. His research interests include cybersecurity and trustworthy AI for cyber-physical systems, with an emphasis on detecting, reasoning about, and mitigating malicious behaviors in complex systems. He is a Full Member of Sigma Xi, The Scientific Research Honor Society.
\end{IEEEbiography}

\vskip -2\baselineskip plus -1fil

\begin{IEEEbiography}
[{\includegraphics[width=1in,height=1.25in,clip,keepaspectratio]{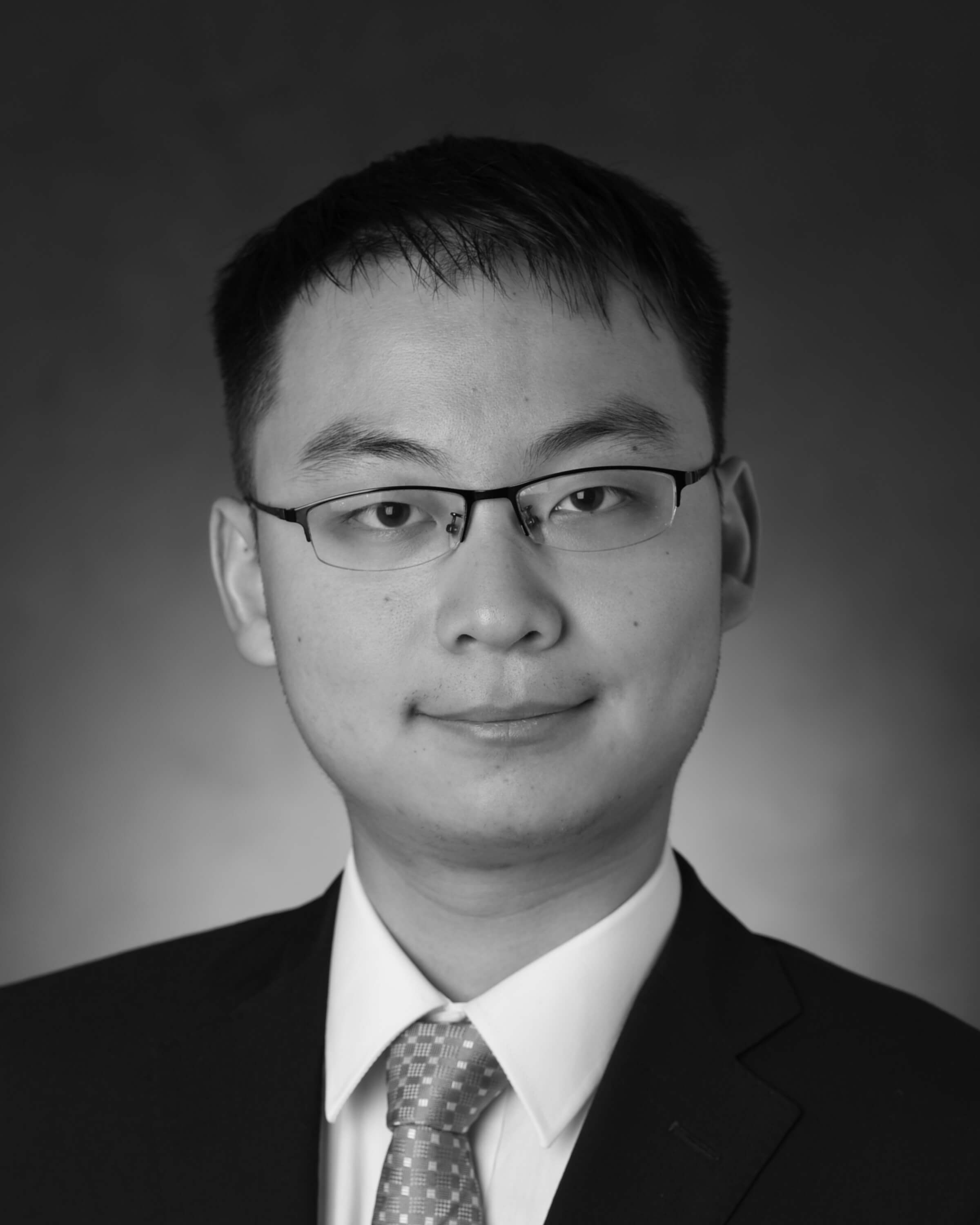}}]{Haizhou Wang}
is a Senior Staff Software Engineer at Palo Alto Networks. He received his PhD. Degree in Informatics at the Pennsylvania State University. His research focus on automated system on file based malware defense and source code level software security.
\end{IEEEbiography}

\vskip -2\baselineskip plus -1fil

\begin{IEEEbiography}
[{\includegraphics[width=1in,height=1.25in,clip,keepaspectratio]{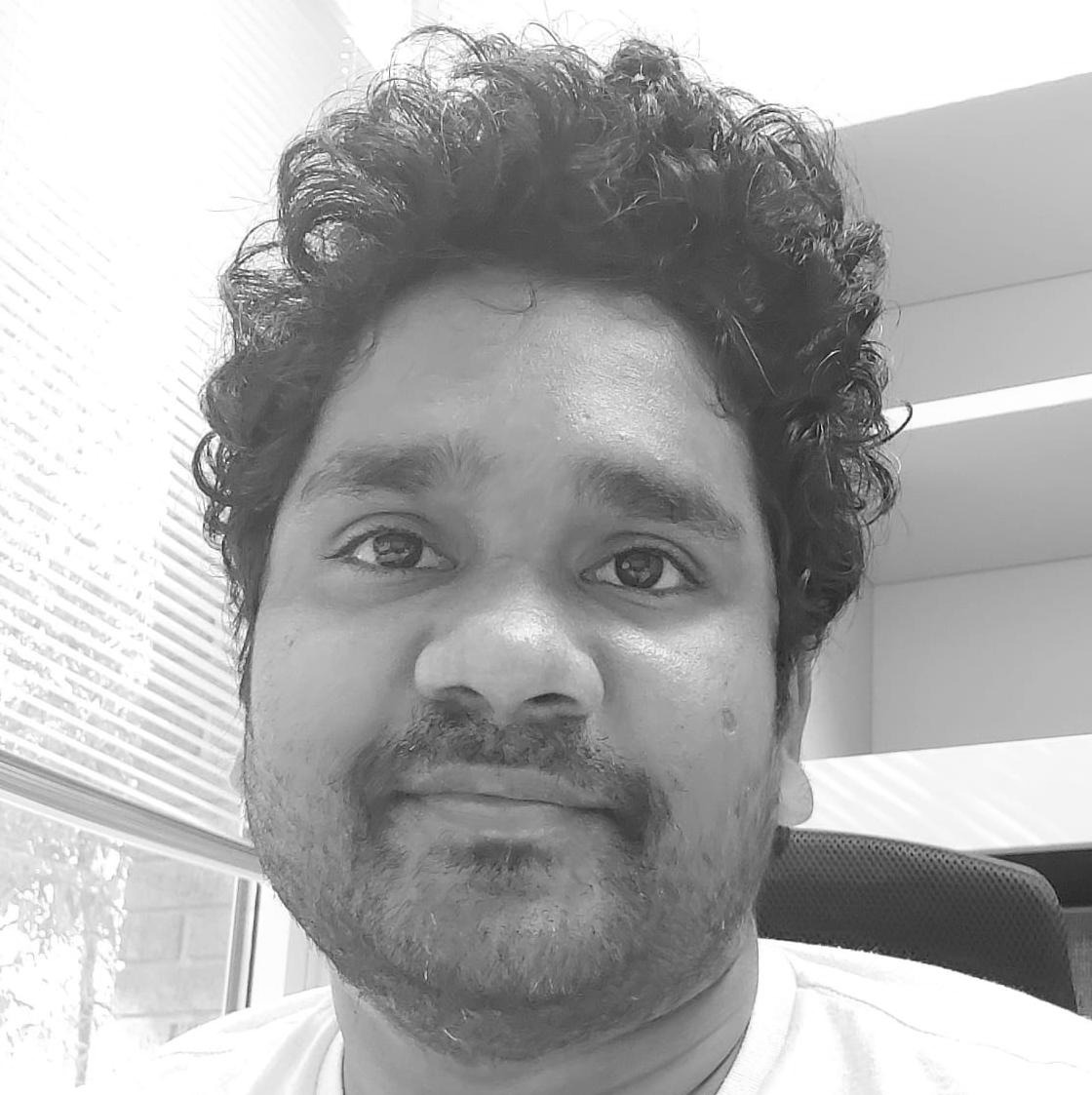}}]{Motakatla Venkateswara Reddy}
received the B.Tech. degree in electrical and electronic engineering and the M.Tech. degree in power electronics and drives from Jawaharlal Nehru Technological University, Kakinada, India, in 2009 and 2013, respectively. He completed Ph.D. degree from the Indian Institute of Technology Ropar, Punjab, India, in 2019 and received POSOCO Power System Award in 2020 for his PhD thesis. Subsequently, he held the position of Postdoctoral Research Associate at Washington State University and Clemson University between 2020-2022. He is currently at a Researcher-III position with the Grid Automation and Control, National Laboratory of the Rockies, Golden, CO, USA. He is contributing to multiple DOE-funded cybersecurity projects and other projects including Sensor-MAP, Federated Architecture for Secure and Transactive DER Management Solutions (FAST-DERMS) and advanced distribution management systems (ADMS) Test Bed use cases. His research interests include AI/ML applications to power system, ADMS/DERMS/VPP, CIM compliant Open-source ADMS platform development, DER OT Cybersecurity, pattern recognition algorithms, power quality issues and power electronics converter applications.
\end{IEEEbiography}

\vskip -2\baselineskip plus -1fil

\begin{IEEEbiography}
[{\includegraphics[width=1in,height=1.25in,clip,keepaspectratio]{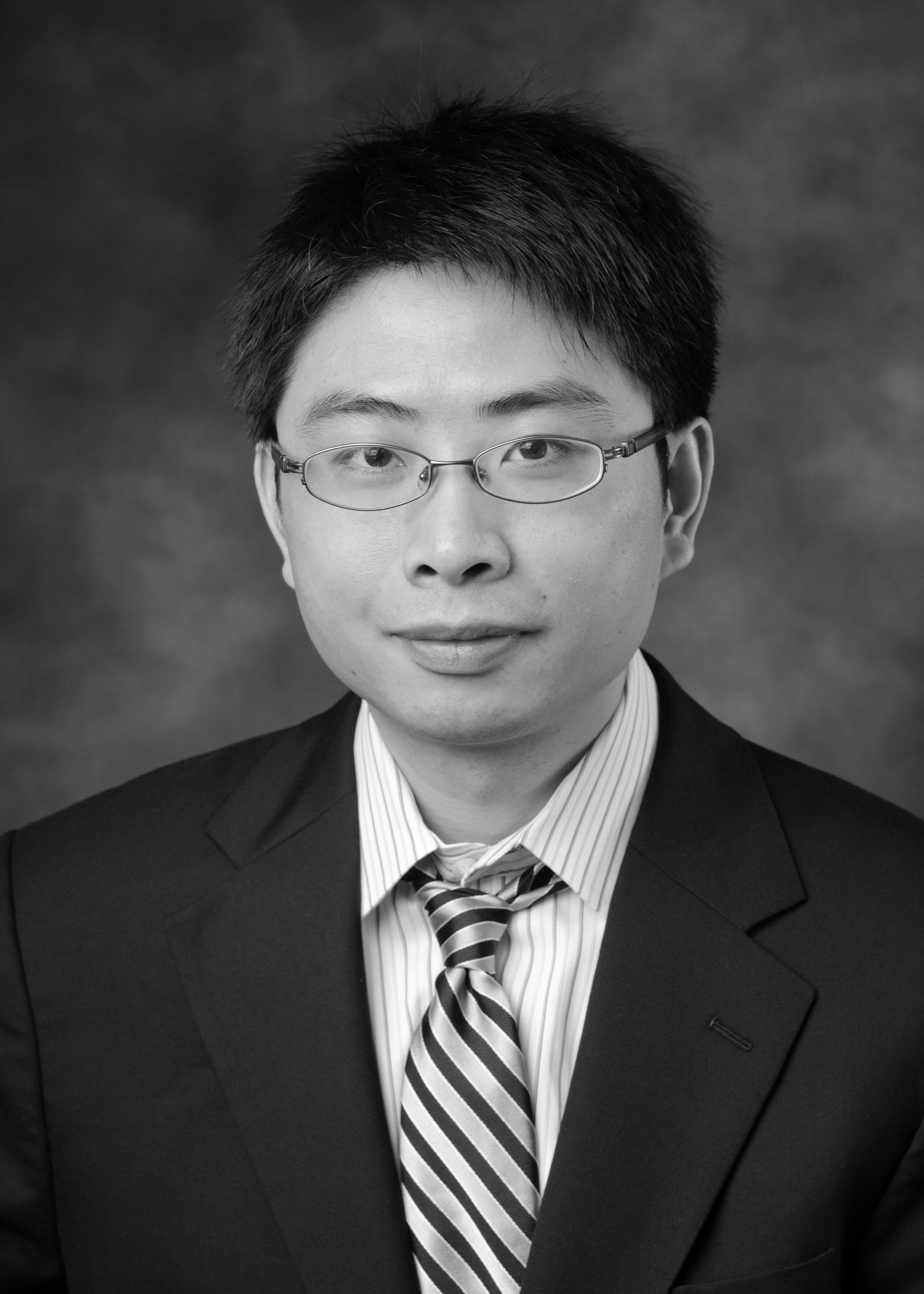}}]{Minghui Zhu}
is a Professor in the School of Electrical Engineering and Computer Science at the Pennsylvania State University. Prior to joining Penn State in 2013, he was a postdoctoral associate in the Laboratory for Information and Decision Systems at the Massachusetts Institute of Technology. He received Ph.D. in Engineering Science (Mechanical Engineering) from the University of California, San Diego in 2011. He was a joint appointee as Senior Engineer in the Optimization and Controls Department at the Pacific Northwest National Laboratory. His research interests lie in distributed control and decision-making of multi-agent networks with applications in robotic networks, security and the smart grid. He is an author of the book “Distributed optimization-based control of multi-agent networks in complex environments” (Springer, 2015). He is a recipient of the Dorothy Quiggle Career Development Professorship in Engineering at Penn State in 2013, the National Science Foundation CAREER award in 2019, the Superior Paper Award of the American Society of Agricultural and Biological Engineers in 2024 and the Penn State Engineering Alumni Society Outstanding Research Award in 2025. He is an associate editor of Automatica, the IEEE Transactions on Automatic Control, the IEEE Open Journal of Control Systems and the IET Cyber-systems and Robotics.
\end{IEEEbiography}

\vskip -2\baselineskip plus -1fil

\begin{IEEEbiography}
[{\includegraphics[width=1in,height=1.25in,clip,keepaspectratio]{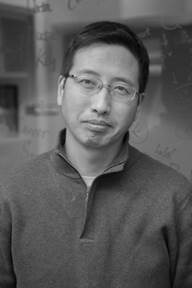}}]{Peng Liu}
is the Raymond G. Tronzo, MD Professor of Cybersecurity, and Director of the Cyber Security Lab at Penn State University.   His research interests are in all areas of computer security.  He has published over 360 technical papers. He has served on over 100 program committees and reviewed papers for numerous journals. He is currently the Co-Editor-in-Chief of Journal of Computer Security. 
\end{IEEEbiography}

\end{document}